\documentclass[aps,prd,preprint,groupedaddress,showpacs,nofootinbib,
floatfix]{revtex4-1}
\pdfoutput=1

\usepackage{amsmath}
\usepackage{amssymb}
\usepackage{mathtools}
\usepackage{graphics}

\usepackage[caption=false]{subfig}

\begin{document}

\title{Late-time quantum backreaction from inflationary fluctuations 
       of a non-minimally coupled massless scalar}

\author{D.~Glavan}
\email[]{d.glavan@uu.nl}

\author{T.~Prokopec}
\email[]{t.prokopec@uu.nl}

\author{D.~C.~van der Woude}
\email[]{d.c.vanderwoude@students.uu.nl}

\affiliation{Institute for Theoretical Physics and Spinoza Institute, Utrecht University,\\
Postbus 80.195, 3508 TD Utrecht, The Netherlands}

\date{\today}

\begin{abstract}
We consider the late time one-loop quantum
backreaction from inflationary fluctuations
of a non-minimally coupled, massless scalar field.
The scalar is assumed to be a spectator
field in an inflationary model with a constant
principal slow roll $\epsilon$ parameter.
We regulate the infrared by matching
onto a pre-inflationary radiation era.
We find a large late time backreaction when the nonminimal coupling $\xi$ is negative
(in which case the scalar exhibits a negative mass term during inflation).
The one-loop quantum backreaction becomes significant today for moderately small
non-minimal couplings, $\xi\sim -1/20$, and it changes sign (from negative to positive)
at a recent epoch when inflation lasts not much longer than what is minimally required,
$N \gtrsim 66$. Since currently we do not have a way of treating the
classical fluid and the quantum backreaction
in a self-consistent manner, we cannot say decidely whether
the backreaction from inflationary quantum fluctuations of a non-minimally coupled scalar can
mimic dark energy.

\end{abstract}


\maketitle

\section{Introduction}
\label{sec:Introduction}

 In Ref.~\cite{Glavan:2013mra} we have studied the late
 time one-loop quantum backreaction
of a minimally coupled scalar field and gravitons from the 
amplified inflationary vacuum
fluctuations on de Sitter space and showed that at late times inflationary
fluctuations scale as the (matter) background and contribute only a tiny amount
($\sim 10^{-13}$) of the total energy density, thereby 
constituting a negligible backreaction
on the background evolution. 
This result was subsequently confirmed in \cite{Aoki:2014ita}, 
where it was also found that the existence of additional pre-inflationary 
eras and transitions between them can increease the amplitude of the
backreaction, but not change its scaling.
Additionally in~\cite{Glavan:2013mra}, by allowing the Hubble parameter to
exhibit adiabatic dependence on time, we have made 
an estimate of the quantum backreaction
from scalar cosmological perturbations, and arrived 
at the conclusion that it is also negligible
($\sim 10^{-12}$). The ratio of the tensor and 
scalar backreactions is approximately equal to
the inflationary $r$ parameter, which has recently been measured to be
about $0.2$~\cite{Ade:2014xna}. These considerations 
refute the claims made some time
ago~\cite{Kolb:2005da, Barausse:2005nf} that
inflationary vacuum fluctuations can play the role of the dark energy today
(albeit already soon thereafter that was questioned in Ref.~\cite{Hirata:2005ei}).
However, in Ref.~\cite{Ringeval:2010hf} it was argued 
that quantum backreaction from inflationary perturbations
of a light, but massive, scalar field can mimic dark energy.

 Even though the backreaction from massless minimally coupled scalars
and from cosmological perturbations is tiny, it is interesting
that it scales precisely as the background, constituting a tiny contribution
to dark matter.~\footnote{It would be of interest to check whether 
the same can be said about the
quantum backreaction from the inflationary perturbations of a light massive 
scalar field whose mass is smaller than the Hubble rate today
($m<H_0$), see Ref.~\cite{Ringeval:2010hf}. A na\^ive expectation 
(which would have to be confirmed
or rebutted by a detailed analysis) is that
the energy density from light inflationary fluctuations scales during inflation as,
$\rho_q\sim H^4_I$ (where $H_I$ is the Hubble rate in inflation), 
while during radiation and matter eras it scales as $\propto 1/a^4$ and 
$\propto 1/a^3$,respectively.}
This result has inspired us to consider quantum 
backreaction from inflationary fluctuations
of a non-minimally coupled spectator scalar field, because for
a particular choice of the non-minimal coupling 
(which is in our convention negative),
the spectrum of amplified vacuum fluctuations can 
be dramatically red-tilted, thereby enhancing
the late time quantum backreaction, {\it cf.} Ref.~\cite{Janssen:2009nz}.
At least that was our hope.
In this paper we confirm this expectation: 
for a sufficiently large negative non-minimal
coupling $\xi$ late time backreaction becomes 
indeed strong. We find that for $\xi<0$
the backreaction grows during inflation, and for 
a minimal inflation with a number of
e-folds $N\simeq 66$ (as required by observations),
 the backreaction is negative and for $\xi\lesssim -0.057$
it exceeds the background density at the end of inflation. 
This opens the possibility to use
quantum fluctuations of a non-minimally coupled scalar 
to terminate inflation, thus providing
a graceful exit from inflation. A proper study of this 
would require a self-consistent solution of
the semiclassical background equation, which is beyond the scope of this paper.
At late times during matter era we find that the backreaction is negative for
$\xi<0$ and it grows with respect to
the background if $\xi<-1/3$. However, for these values
the backreaction is already too large during inflation, and hence
it ought to be discarded. As an interesting case we
present a universe with an inflationary period of $N\simeq 69$ and with
$\xi\simeq -0.052$. In this case the backreaction 
is subdominant during inflation and radiation era and
it begins negative in matter era, but then it 
exhibits a transient and becomes positive at recent times,
ending up decaying in the future. This case is 
interesting since it could be potentially used to
mimic dark energy. However, a more sophisticated 
self-consistent analysis of the background
evolution that includes the backreaction
in the background is needed to properly answer 
that question, and we postpone this analysis to future work.

 The question of quantum backreaction of 
cosmological perturbations and other
quantum fields during inflation is an old subject, 
and has been extensively studied in literature.
A particular attention was devoted to the one 
loop quantum backreaction of scalar and tensor cosmological
perturbations during inflation~\cite{Abramo:2001dc,Abramo:2001db,
Abramo:1998hj,Abramo:1998hi,
Geshnizjani:2002wp,Geshnizjani:2003cn,Abramo:2001dd,Janssen:2008dw}.
A consensus has been reached that a reasonable local observer will observe the local
expansion rate~\cite{Geshnizjani:2002wp,Geshnizjani:2003cn,Abramo:2001dd},
while a non-local observer would observe a backreaction 
accumulated along the past light
cone~\cite{Abramo:2001dc,Abramo:2001db,Abramo:1998hj,Abramo:1998hi}.
In both cases the quantum backreaction vanishes 
at the one loop order. There are however observers for
which the one loop quantum backreaction does not vanish
~\cite{Marozzi:2013uva,Marozzi:2011zb,Marozzi:2012tp},
emphasising the observer's importance
when deciding on the quantum backreaction during inflation.

Because of the strongly non-conformal nature 
f the coupling of scalar and tensor perturbations
to an expanding background, these fields are naturally
believed to yield a particularly strong quantum backreaction,
it is worth investigating the quantum backreaction 
f other quantum fields, which also
(at the classical or quantum level) break conformal invariance.
Thus, the two-loop quantum backreaction on de Sitter background was considered in
Refs.~\cite{Prokopec:2006ue,Prokopec:2008gw}, where it was found that
the (negative) backreaction (on the background energy density) is of the order,
$\sim -10^{-2}\alpha_e H^4\ln(a)$.
On the other hand, its non-perturbative generalization based on stochastic
inflation~\cite{Prokopec:2007ak} yields at late times in inflation
a parametrically larger (negative) backreaction of the order, $\sim -10^{-2}H^4$,
that is of the order $-G_N\Lambda$ times the background density.
It is remarkable that this result is independent 
on the electromagnetic coupling constant $\alpha_e$,
and hence it is fully non-perturbative. Other 
results of interest include quantum backreaction in
Yukawa theory~\cite{Miao:2006pn}, in which 
the fermion loops give a negative contribution that grows
in magnitude as the scalar field vacuum expectation 
value, thus destabilizing de Sitter.
Finally, it is worth mentioning that the quantum 
scalar perturbations on de Sitter space
are strong enough to restore an $O(N)$
symmetry~\cite{Lazzari:2013boa,Serreau:2013eoa,Prokopec:2011ms,Serreau:2011fu,Janssen:2009pb,
Starobinsky:1994bd,Vilenkin:1982wt}, 
and thus induce a large quantum backreaction~\cite{Lazzari:2013boa},
albeit it is at the moment not clear whether this effect
can have significant late time impact on the evolution of the Universe
and on cosmological perturbations.

\begin{figure}[h]

\begin{turnpage}

\includegraphics{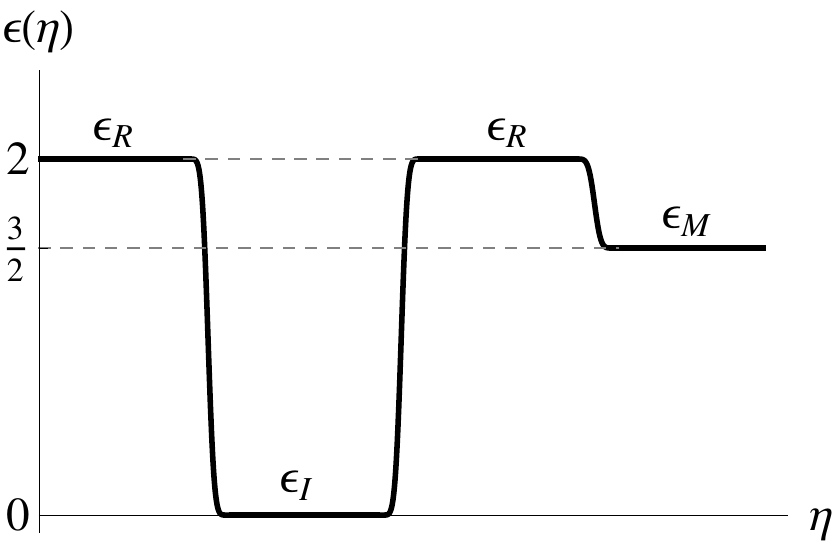}
\caption{The evolution of the slow-roll parameter $\epsilon(\eta)$ assumed
in this paper.} 
\label{expansion history}

\end{turnpage}

\end{figure}

\section{FLRW background}
\label{sec:FLRW background}

The metric of a $D$-dimensional spatially flat
Friedmann-Lema\^{i}tre-Robertson-Walker (FLRW) space-time,
in conformal time coordinates, is
\begin{equation}
g_{\mu\nu}(x) = a^2(\eta) \eta_{\mu\nu},
\end{equation}
where $\eta_{\mu\nu}=\mathrm{diag}(-1,1,1,\dots,1)$ is
the $D$-dimensional Minkowski metric, and $a(\eta)$ is the scale
factor. We work in units $\hbar=c=1$ unless otherwise stated. 
The conformal time is
related to the physical (cosmological) time $t$ as $dt=a(\eta)d\eta$.
The conventions we use for the geometric quantities are
$\Gamma^\alpha_{\mu\nu}=
\frac{1}{2}g^{\alpha\beta}(\partial_\mu g_{\nu\beta}
+\partial_\nu g_{\mu\beta} -\partial_\beta g_{\mu\nu})$
for Christoffel symbols,
${R^\alpha}_{\mu\beta\nu} = \partial_\beta \Gamma^\alpha_{\mu\nu}
- \partial_\nu \Gamma^\alpha_{\mu\beta}
+ \Gamma^{\alpha}_{\beta\rho}\Gamma^\rho_{\mu\nu}
- \Gamma^\alpha_{\nu\rho}\Gamma^\rho_{\mu\beta}$
for the Riemann tensor,
$R_{\mu\nu}={R^\alpha}_{\mu\alpha\nu}$
for the Ricci tensor, and
$R={R^\mu}_{\mu}$ is the Ricci scalar.

The source for the dynamics of this space-time
(the dynamics of the scale factor) is the matter
content of the Universe, which we take to be
in the form of ideal fluids. Ideal fluids are described
by their energy density $\rho$ and pressure $p$.
The cosmologically relevant fluids satisfy
the linear equation of state $p=w\rho$, where
$w_{\scriptscriptstyle{M}}=0$ for non-relativistic
matter (dust), $w_{\scriptscriptstyle{R}}=1/3$
for radiation (ultra-relativistic matter), and
$w_{\scriptscriptstyle{\Lambda}}=-1$ for
cosmological constant.

The dynamics of the FLRW space-time is dictated by the two Friedmann equations,
\begin{align}
& \left( \frac{\mathcal{H}}{a} \right)^2
	= \frac{8\pi G_{\!\scriptscriptstyle{N}}}{3} \sum_i \rho_i , \label{friedmann1}
\\
& \frac{\mathcal{H}'-\mathcal{H}^2}{a^2}
	= - 4\pi G_{\!\scriptscriptstyle{N}} \sum_i 
		\left(\rho_i + p_i \right) ,
\end{align}
where we have introduced a conformal Hubble rate,
\begin{equation}
\mathcal{H}= \frac{a'}{a},
\end{equation}
the prime denoting differentiation with respect to conformal time.
It is related to the physical Hubble rate as $H=\mathcal{H}a^{-1}$.
Also, $G_{\!\scriptscriptstyle{N}}$ is
the Newton's constant and $c$ the speed of light.

The background ideal fluids satisfy each independently the
covariant conservation law,
\begin{equation} \label{conservation eq}
\rho' + 3\mathcal{H} \left(\rho + p \right) = 0 .
\end{equation}
When the equation of state is linear, $p= w \rho$, this conservation law is readily integrated
to yield the scaling of energy density and pressure with the scale factor,
\begin{equation}
\rho = \frac{p}{w} = \rho_0 \left( \frac{a_0}{a} \right)^{3(1+w)},
\end{equation}
where $\rho(\eta_0)=\rho_0$ and $a(\eta_0)=a_0$, and $w$ is the 
equation of state parameter.

When expansion of the Universe is dominated by a single fluid component
(so that other fluids can be neglected), 
the Friedmann equations can easily be integrated to solve for the scale factor,
\begin{equation}
a(\eta) = a_0 \Big{[} 1 + (\epsilon\!-\!1)\mathcal{H}_0(\eta\!-\!\eta_0)
	\Big{]}^{\frac{1}{\epsilon-1}},
\end{equation}
and the Hubble rate,
\begin{equation} \label{HinA}
\mathcal{H} = \mathcal{H}_0 \left( \frac{a_0}{a} \right)^{\epsilon-1} ,
\end{equation}
where $\mathcal{H}(\eta_0)=\mathcal{H}_0$, and $\epsilon$ is the principal
slow-roll parameter (the terminology borrowed from inflation, but here
it does not have to be small), generally defined as
\begin{equation}
\epsilon = -\frac{\dot H}{H^2} = 1 \!-\! \frac{\mathcal{H}'}{\mathcal{H}^2} .
\end{equation}
For spatially flat FLRW space-times dominated by one ideal fluid $\epsilon$ is constant,
and depends only on the fluid equation of state parameter,
\begin{equation}
\epsilon = \frac{3}{2}(1+w) .
\end{equation}
The $\epsilon$ parameter is connected to the deceleration parameter
as $q=\epsilon-1$, which
is usually used to quantify whether and how much
the expansion of the Universe is speeding up or slowing down.
That is obvious from its definition
written in terms of physical time,
\begin{equation}
q = \epsilon \!-\! 1 = \!-\! \frac{\ddot{a}a}{\dot{a}^2} ,
\end{equation}
which tells us that the Universe is decelerating when $\epsilon>1$, and it is
accelerating when $\epsilon<1$. The relevant values of this parameter are
$\epsilon_{\scriptscriptstyle{M}}=3/2$ for non-relativistic matter,
$\epsilon_{\scriptscriptstyle{R}}=2$ for radiation and
$0<\epsilon_{\scriptscriptstyle{I}}\ll1$ for inflation (close to that of the
cosmological constant $\epsilon_{\scriptscriptstyle{\Lambda}}=0$).

The picture one should keep in mind when it comes to the
expansion history in this paper is shown in Figure
\ref{expansion history}. The transition times
between periods $\tau_n$ are assumed to be fast,
$\tau_n\ll 1/\mathcal{H}_n$. The initial radiation-dominated
period is assumed since it serves as a universal IR regulator~\cite{Janssen:2009nz}
for the initial scalar field state.

\section{Non-minimally coupled scalar field}
\label{sec:Non-minimally coupled scalar field}

The action for a non-minimally coupled massless scalar field is given by
\begin{equation}
S = \int d^D\!x\, \sqrt{-g}
	\left(\! -\frac{1}{2} \, g^{\mu\nu} \partial_\mu\phi \, \partial_\nu\phi
	-\frac{1}{2}\xi R \phi^2 \right),
\end{equation}
where the dimensionless parameter $\xi$ is the non-minimal coupling
and $g=\mathrm{det}(g_{\mu\nu})$.
Note that the sign convention for $\xi$ we use is the opposite from the one used in the
literature on Higgs inflation, such that 
$\xi=\xi_c=(D-2)/[4(D-1)]$ is the conformal coupling in our convention.
In order to quantize this field on a FLRW background
we first define a canonically conjugate momentum
\begin{equation}
\pi(x) = \frac{\delta S}{\delta \phi'(x)} = a^{D-2}(\eta) \phi'(x) \ ,
\end{equation}
then promote $\phi$ and $\pi$ to operators and impose
canonical commutation relations (we work in the Heisenberg picture),
\begin{equation}
\big[\hat{\phi}(\eta,\boldsymbol{x}) , \hat{\pi}(\eta,\boldsymbol{x}')\big] =
	i \delta^{D-1}(\boldsymbol{x}\!-\!\boldsymbol{x}') , \qquad
	\big[\hat{\phi}(\eta,\boldsymbol{x}) ,
		\hat{\phi}(\eta,\boldsymbol{x}')\big] = 0
	= \big[\hat{\pi}(\eta,\boldsymbol{x}) , \hat{\pi}(\eta,\boldsymbol{x}')\big] .
\end{equation}
The Heisenberg equations then give us the equation of motion for the field operator,
\begin{equation}
\hat{\phi}'' + (D-2) \mathcal{H}\hat{\phi}' - \nabla^2\hat{\phi}
	+\xi(D-1)\Big{(} 2\mathcal{H}' + (D-2)\mathcal{H}^2 \Big{)} \hat{\phi} = 0
\,,
\end{equation}
and we have assumed that gravity is non-dynamical (for a discussion of the one-loop 
graviton contribution see Ref.~\cite{Glavan:2013mra}).
Due to the spatial translation and rotation invariance of the background, it 
is convenient to decompose the field operator into Fourier modes,
\begin{equation}
\hat{\phi}(\eta,\boldsymbol{x}) =
	a^{\frac{2-D}{2}} \int \frac{d^{D-1}k}{(2\pi)^{D-1}}
	\Big{(} e^{i\boldsymbol{k}\cdot\boldsymbol{x}} U(k,\eta)
		\hat{b}(\boldsymbol{k})
	+ e^{-i\boldsymbol{k}\cdot\boldsymbol{x}} U^*(k,\eta)
		\hat{b}^{\dag}(\boldsymbol{k})
	\Big{)},
\end{equation}
where the mode function $U(k,\eta)$ is assumed to depend
only on the modulus of the momentum
$k=\| \boldsymbol{k} \|$ (corresponding to an isotropic state).
Its equation of motion is
\begin{equation} \label{EOM}
U''(k,\eta) + \Big{(} k^2 + f(\eta) \Big{)}U(k,\eta) = 0,
\end{equation}
where
\begin{equation} \label{f}
f(\eta) = -\frac{1}{4}\Big{(}D-2-4\xi(D-1)\Big{)}
	 \Big{(}2\mathcal{H}'+(D-2)\mathcal{H}^2\Big{)}.
\end{equation}
We require the annihilation and creation operators to satisfy the
following commutation relations,
\begin{equation}
\left[\hat{b}(\boldsymbol{k}), \hat{b}^{\dag}(\boldsymbol{k}')\right]
	= (2\pi)^{D-1} \delta^{D-1}(\boldsymbol{k}\!-\!\boldsymbol{k}'), \qquad
	\left[\hat{b}(\boldsymbol{k}), \hat{b}(\boldsymbol{k}')\right] = 0
	= \left[\hat{b}^{\dag}(\boldsymbol{k}), \hat{b}^{\dag}
		(\boldsymbol{k}')\right] ,
\end{equation}
which then fixes the Wronskian normalization of the mode function
\begin{equation} \label{Wronskian}
\mathcal{W} \left[U(k,\eta),U^*(k,\eta) \right] =
	U(k,\eta) \frac{\partial}{\partial\eta} U^*(k,\eta)
	\!-\! U^*(k,\eta) \frac{\partial}{\partial\eta} U(k,\eta) = i.
\end{equation}
The vacuum state $|\Omega\rangle$ is defined to be annihilated by the annihilation operators,
\begin{equation}
\hat{b}(\boldsymbol{k}) |\Omega\rangle = 0, \qquad \forall \boldsymbol{k},
\end{equation}
and the entire Fock space is constructed 
by acting on it by the creation operators $\hat{b}^{\dag}(\boldsymbol{k})$.
It is completely determined once we have specified what is the mode
function and its derivative at some time. The initial state we pick is
the Bunch-Davies (B-D) vacuum $u(k,\eta)$, which is the one that reduces to the flat
space positive-frequency mode function in the UV,
\begin{equation} \label{BD UV}
U(k,\eta)\stackrel{\rm BD}{\longrightarrow}
u(k,\eta) \;\overset{k\rightarrow\infty}{\sim}\; \frac{e^{-ik\eta}}{\sqrt{2k}} ,
\end{equation}
since the high momentum modes are insensitive to curvature, as can be seen
{\it e.g.} from the UV adiabatic expansion.
Physically, this simply means that exciting high momentum modes costs a lot of energy, and 
is therefore highly suppressed.
On constant $\epsilon$
backgrounds (with which we will be concerned here) it is possible to
extend definition (\ref{BD UV}) to a global Bunch-Davies state,
which will be given explicitly in section
\ref{sec:Mode function}
(for a more detailed discussion on
this see~Ref.\cite{Glavan:2013mra}).

\section{Energy-momentum tensor}
\label{sec:Energy-momentum tensor}

The energy-momentum tensor operator of a massless non-minimally coupled
scalar field is defined to be
\begin{equation}
\hat{T}_{\mu\nu}
	= \left. \frac{-2}{\sqrt{-g}} \frac{\delta S}{\delta g^{\mu\nu}}
		\right|_{\phi\rightarrow\hat{\phi}}
	= \partial_\mu\hat{\phi} \, \partial_\nu\hat{\phi}
	- \frac{1}{2}g_{\mu\nu} g^{\alpha\beta}
		\partial_\alpha\hat{\phi} \, \partial_\beta\hat{\phi}
	+ \xi \Big{(} G_{\mu\nu} - \nabla_\mu\nabla_\nu
	+g_{\mu\nu}\square \Big{)} \hat{\phi}^2,
\end{equation}
where $G_{\mu\nu}=R_{\mu\nu}-\frac{1}{2}g_{\mu\nu}R$ is the
Einstein tensor, $\nabla$ denotes the covariant derivative, and
$\square = g^{\mu\nu}\nabla_\mu\nabla_\nu$.
The expectation value of this operator on FLRW with respect to the
isotropic state is diagonal, and it can be written as integrals in
Fourier space over squares of the mode function,
\begin{align}
\langle\Omega| \hat{T}_{00}|\Omega\rangle ={}&
	\frac{a^{2-D}}{(4\pi)^{\frac{D-1}{2}}\Gamma\left( \frac{D-1}{2} \right)}
	\int\limits_{0}^{\infty} dk\, k^{D-2} \Bigg{[} 2k^2|U|^2
	- \frac{1}{2}\Big{(}D-2-4\xi(D-1)\Big{)} \mathcal{H}' |U|^2
\nonumber \\
& 	\qquad - \frac{1}{2}\Big{(}D-2-4\xi(D-1)\Big{)} \mathcal{H}
	\frac{\partial}{\partial\eta}|U|^2
	 + \frac{1}{2} \frac{\partial^2}{\partial\eta^2}|U|^2 \Bigg{]}, \label{rho def}
\\
\langle\Omega| \hat{T}_{ij}|\Omega\rangle ={}&
	\frac{\delta_{ij}a^{2-D}}{(4\pi)^{\frac{D-1}{2}}\Gamma
	\left( \frac{D-1}{2} \right)}
	\int\limits_{0}^{\infty} dk\, k^{D-2} \Bigg{[} \frac{2k^2}{D-1}|U|^2
	- \frac{1}{2} \Big{(}D-2-4\xi(D-1)\Big{)} \mathcal{H}' |U|^2
\nonumber \\
& 	\qquad - \frac{1}{2}\Big{(}D-2-4\xi(D-1)\Big{)} \mathcal{H}
	\frac{\partial}{\partial\eta}|U|^2
	 + \frac{1}{2}(1-4\xi) \frac{\partial^2}{\partial\eta^2}|U|^2 \Bigg{]},
\label{p def}
\end{align}
where we have exploited the fact that the mode function depends only on
the modulus of the momentum and performed the angular integrations
right away. In addition, we have used the equation of motion~(\ref{EOM}) to rewrite
$|U'|^2 = (k^2+f)|U|^2 + \frac{1}{2}\frac{\partial^2}{\partial\eta^2}|U|^2$.
The two quantities above are divergent in $D=4$
(quartically, quadratically and logarithmically), hence
they need to be regularized and renormalized.
The regularization used here is the dimensional regularization 
(that is why we kept all the expressions in arbitrary $D$ dimensions) 
which automatically removes the quartic and quadratic
divergences. As shown in Appendix~\ref{sec:Renormalization}, 
the logarithmic divergences are absorbed into higher
derivative counter-terms, after which the limit
$D\rightarrow4$ can be taken
and finite physical values associated with the two integrals above.
From now on we express
this expectation value of the energy-momentum tensor in terms of the energy
density and pressure (of the quantum fluid),
\begin{equation}
\rho_q = \frac{1}{a^2} \langle\Omega| \hat{T}_{00} |\Omega\rangle, \qquad
	\delta_{ij} p_q = \frac{1}{a^2} \langle\Omega| \hat{T}_{ij} |\Omega\rangle.
\label{rhoq and pq}
\end{equation}
The ratio of these two quantities is the equation of state parameter for the quantum fluid,
\begin{equation}
w_q = \frac{p_q}{\rho_q},
\end{equation}
which is sometimes a convenient quantity to use when comparing it to the classical fluid
which drives the expansion. This energy density and pressure have to satisfy
the conservation equation
\begin{equation}
\rho_q' + 3\mathcal{H}(\rho_q + p_q) = 0,
\end{equation}
which also serves as an independent check of the calculation.

\section{Evolution of the mode function and Bogolyubov coefficients}
\label{sec:Mode function}

Since we will be concerned with the scalar field evolving on FLRW with a few periods
of constant $\epsilon$ expansion (as shown in Figure~\ref{expansion history}),
 it is convenient to examine what the Bunch-Davies
state is on those periods. On accelerating periods ($\epsilon<1$) 
the B-D mode function is~\cite{Janssen:2009nz}:
\begin{equation} \label{u accelerating}
u(k,\eta) = \sqrt{\frac{\pi}{4(1-\epsilon)\mathcal{H}}} \
	H_\nu^{(1)}\left( \frac{k}{(1-\epsilon)\mathcal{H}} \right),
\end{equation}
and on decelerating periods ($\epsilon>1$) we take it to be
\begin{equation} \label{u deceleration}
u(k,\eta) = \sqrt{\frac{\pi}{4(\epsilon-1)\mathcal{H}}} \
	H_\nu^{(2)}\left( \frac{k}{(\epsilon-1)\mathcal{H}} \right),
\end{equation}
where in both cases
\begin{equation}
\nu^2 = \frac{1}{4}+ 
	\frac{D-2\epsilon}{4(1-\epsilon)^2} \Big{(} D-2-4\xi(D-1) \Big{)} ,
\label{nu D}
\end{equation}
and $H_\nu^{(1)}$ and $H_\nu^{(2)}$ are the first and second Hankel functions,
respectively.\footnote{Our choice for the B-D 
mode function on decelerating periods might
seem odd, since usually (\ref{u accelerating}) is a B-D mode function on any
constant $\epsilon$ period. But, since we assume that $\nu$ is real (in order to
have appreciable particle production during inflation), 
$H_\nu^{(1)}(-x)$ differs from $H_\nu^{(2)}(x)$
by an irrelevant phase factor, our choice corresponds to picking a particular phase
which does not affect physical quantities. This way, 
the arguments of the Hankel functions
are always positive, which we find more convenient to work with.}

Even if the field starts in a B-D state during some period of 
constant $\epsilon=\epsilon_0$,
if $\epsilon$ evolves, so will the state 
and when $\epsilon$ eventually settles to a new constant value $\epsilon=\epsilon_1\neq \epsilon_0$,
the state will differ from the B-D state with $\epsilon=\epsilon_1$
(characterized by $u(k,\eta)$). The scalar
mode function can then be written as, 
\begin{equation}
U(k,\eta) = \alpha(k) u(k,\eta) + \beta(k) u^*(k,\eta),
\end{equation}
where $\alpha$ and $\beta$ are the Bogolyubov coefficients which, 
because of the space-time homogeneity,
depend on the modulus of the momentum, and do not mix different modes
since the equation of the motion for the field is linear. 
This is nothing other than representing the full solution to the
equation of motion (\ref{EOM}) in the basis of B-D solutions.
The actual form of the Bogolyubov coefficients depends on the exact evolution
of the background from one period to the other. The canonical quantization
puts the following constraint on these coefficients,
\begin{equation} \label{unitarity}
|\alpha(k)|^2 - |\beta(k)|^2 = 1 ,
\end{equation}
which is dictated by the Wronskian normalization (\ref{Wronskian}).
The evolution of the mode function on a smooth 
FLRW background also puts constraints
on these coefficients in the UV. Namely, they
 have to reduce to $\alpha\rightarrow1$ and
$\beta\rightarrow0$ as $k\rightarrow\infty$ faster than any power of $1/k$ (see
Appendix~\ref{sec:Renormalization}).

Before dealing with the specific structure of the Bogolyubov coefficients, we will
first define a convenient way to split the expectation value for the 
energy-momentum tensor.
On a given $n$-th period of constant $\epsilon_n$ the mode function will be given as
\begin{equation}
U_n(k,\eta) = \alpha_n(k) u_n(k,\eta) + \beta_n(k) u_n^*(k,\eta),
\end{equation}
where $u_n(k,\eta)$ is the Bunch-Davies mode function on a given $n$th period.
Since in the integrals (\ref{rho def}) and (\ref{p def}) only $|U_n|^2$ is needed,
we find it convenient to write it out in terms of B-D mode function as
\begin{equation}
|U_n|^2 = |u_n|^2 + \Big{[} 2|\beta_n|^2|u_n|^2
	+ \alpha_n\beta_n^*u_n^2 + \alpha_n^*\beta_nu_n^{*2} \Big{]} ,
\label{Un squared}
\end{equation}
where (\ref{unitarity}) was used.
To facilitate our analysis, it is convenient to define the energy density and pressure
functionals,
\begin{align}
\widetilde{\rho}_q[Z] ={}& \frac{a^{-D}}
		{(4\pi)^{\frac{D-1}{2}} \Gamma\left( \frac{D-1}{2} \right)}
	\int\limits_{k_0}^{\infty}\! dk\, k^{D-2} \Bigg{[} 2k^2 Z(k,\eta) 
	+ \frac{1}{2} \Big{(} D-2-4\xi(D-1) \Big{)} (\epsilon_n-1)
		\mathcal{H}^2 Z(k,\eta)
\nonumber \\
&	\qquad - \frac{1}{2}\Big{(} D-2-4\xi(D-1) \Big{)}\mathcal{H}
		\frac{\partial}{\partial\eta}Z(k,\eta) 
	+ \frac{1}{2}\frac{\partial^2}{\partial\eta^2}Z(k,\eta) \Bigg{]} ,
\\
\widetilde{p}_q[Z] ={}& \frac{a^{-D}}
		{(4\pi)^{\frac{D-1}{2}} \Gamma\left( \frac{D-1}{2} \right)}
	\int\limits_{k_0}^{\infty}\! dk\, k^{D-2} \Bigg{[} \frac{2k^2}{D-1} Z(k,\eta) 
	+ \frac{1}{2} \Big{(} D-2-4\xi(D-1) \Big{)} (\epsilon_n-1)
		\mathcal{H}^2 Z(k,\eta)
\nonumber \\
&	\qquad - \frac{1}{2}\Big{(} D-2-4\xi(D-1) \Big{)}\mathcal{H}
		\frac{\partial}{\partial\eta}Z(k,\eta) 
	+ \frac{1}{2}(1-4\xi)\frac{\partial^2}{\partial\eta^2}Z(k,\eta) \Bigg{]} ,
\end{align}
Where then the full energy density and pressure are sums of two contributions,
one with $Z=Z_{BD}$ and another with $Z=Z_{Bog.}$ (to be defined below).
The IR regulator $k_0$ introduced here does not mean the state is IR divergent
and needs regularization (starting with an initial radiation period takes care of 
that \cite{Ford:1977in}).
It just means that the individual integrals in which we have split the
expectation value might be IR divergent individually. In the final answer,
when they are all added together, the IR divergent terms cancel
and the limit $k_0\rightarrow 0$ can safely be taken.

The first part of the energy density and pressure
~(\ref{rho def}--\ref{rhoq and pq}) corresponds to the
contribution one would get from assuming a B-D vacuum state
on a given period (so in that sense this part sees no transitions),
\begin{align}
\rho_q^{BD} =\widetilde{\rho}_q[|u_n|^2] ,
\qquad p_q^{BD} = \widetilde{p}_q[|u_n|^2].
\end{align}
This B-D part contains all UV divergences of the energy-momentum tensor,
and the regularization and renormalization procedure outlined in Appendix
\ref{sec:Renormalization} applies to this contribution. The
remaining contributions will not contain any UV divergences.
Since all possible $k_0$-dependent terms cancel, on dimensional grounds,
the contribution of the B-D part must be proportional to $\mathcal{H}^4$.
The rest of the time dependence is $a^{-4}\ln(a)$ at best
(Appendix A).
On a matter-dominated period $\mathcal{H}$ falls off as $a^{-1/2}$,
meaning this B-D contribution to the energy density and pressure is
negligible at late times compared to the background quantities, and we will
not consider it in the following.

The remaining contributions to the energy density and pressure involve integrals
over Bogolyubov coefficients as well, involving in particular $\beta(k)$, 
which is non-adiabatically (faster than any power law)
suppressed in the UV, meaning that the integrals can be
evaluated in $D=4$ right away. These contributions are
\begin{align} \label{rho i p Bog}
\rho_q^{Bog.} = \widetilde{\rho}_q [Z_{Bog.}(k,\eta)] , 
\qquad p_q^{Bog.} = \widetilde{p}_q [Z_{Bog.}(k,\eta)] ,
\end{align}
where
\begin{equation}
Z_{Bog.}(k,\eta) = 2|\beta_n(k)|^2|u_n(k,\eta)|^2 + \alpha_n(k)\beta_n^*(k) u_n^2(k,\eta)
	+ \alpha_n^*(k)\beta_n(k) u_n^{*2}(k,\eta) ,
\end{equation}
%
If there are any interesting effects that depend on the expansion history of
the Universe, they must lie in these contributions, since $\alpha_n$ and
$\beta_n$ carry all the information about the expansion. Therefore, we devote
the rest of the paper to calculating this part.

Calculating (\ref{rho i p Bog}) can be reduced to calculating integrals,
\begin{equation} \label{int I}
\mathcal{I}_s = \int_{k_0}^{\infty}\! dk\, k^{2s} Z_{Bog.}(k,\eta) , \qquad s=1,2,
\end{equation}
and then just acting on them with differential operators,
\begin{align}
\rho_q \approx{}& \rho_q^{Bog.} = \frac{1}{4\pi^2a^4} \left\{ 2 \mathcal{I}_2 
	+ \left[ (1-6\xi)(\epsilon_n-1) \mathcal{H}^2 - (1-6\xi)\mathcal{H} \frac{\partial}{\partial\eta}
	+ \frac{1}{2}\frac{\partial^2}{\partial\eta^2} \right] \mathcal{I}_1 \right\} , 
\label{rhoQinI}
\\
p_q \approx{}& p_q^{Bog.} = \frac{1}{4\pi^2a^4} \left\{ \frac{2}{3} \mathcal{I}_2 
	+ \left[ (1-6\xi)(\epsilon_n-1) \mathcal{H}^2 - (1-6\xi)\mathcal{H} \frac{\partial}{\partial\eta}
	+ \frac{1}{2}(1-4\xi)\frac{\partial^2}{\partial\eta^2} \right] \mathcal{I}_1 \right\} .
\label{pQinI}
\end{align}
Isolating the dominant contributions to the integrands in integrals (\ref{int I})
and calculating the integrals comprises the main part of this work.

\section{Dominant contributions to integrals $\mathcal{I}_s$}
\label{sec:Dominant contributions}

In this section we argue where dominant contributions to integrals
(\ref{int I}) should lie.
We do this by approximating the integrand function 
$Z_{Bog.}(k,\eta)$ on different integration intervals in such a 
way that the integration can be performed. Approximations in 
essence consist of performing 
expansions in certain small ratios of scales.

\begin{figure}[h]
\begin{centering}
\includegraphics[width=8cm]{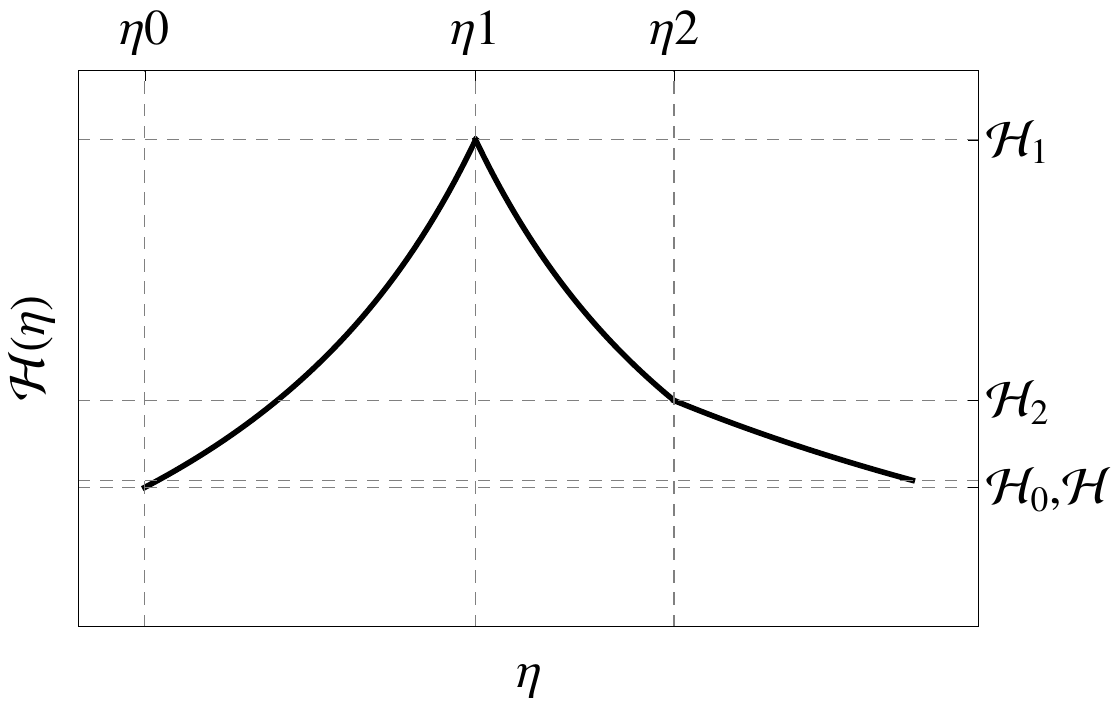}
\end{centering}
\caption{Schematic evolution of the conformal Hubble rate during
different expansion periods. The beginning of inflation is denoted
by $\eta_0$, $\eta_1$ denotes the start of radiation period,
and $\eta_2$ the start of matter period. The hierarchy between scales
today is $\mathcal{H}_1\gg\mathcal{H}_2\gg\mathcal{H}\sim\mathcal{H}_0$.}
\label{Hubble rate}
\end{figure}

The transitions between the periods are assumed fast, so the Hubble rate
at the time of transition is a well defined quantity, since it varies very little
during the time of the transition $\tau\ll1\mathcal{H}$.
Schematically, the evolution of the Hubble rate is depicted in 
Figure \ref{Hubble rate}.

In particular, the hierarchy we are dealing with is $\mathcal{H},\mathcal{H}_0\ll\mathcal{H}_2\ll
\mathcal{H}_1\ll\mathcal{\tau}^{-1}$, 
where $\tau$ is some scale of transition. Therefore, we first split the integration
into the IR and UV part with respect to scales $\mathcal{H},\mathcal{H}_0$.
We do this by introducing a comoving cut-off $\mu$ such that
$\mathcal{H}_0,\mathcal{H}\ll\mu\ll\mathcal{H}_2\ll\mathcal{H}_1$.

\subsection{UV contribution}

First we want to argue that the relevant contribution (if it exists) 
comes from the IR part of the integral $(0,\mu_0)$. Consider the UV integral
\begin{equation}
I_s^{(\mu,\infty)} = \int\limits_{\mu}^{\infty}dk\, k^{2s} Z_{Bog.}(k,\eta)
	= \int\limits_{\mu}^{\infty}dk\, k^{2s} \Big{[} 2|\beta_M|^2 |u_M|^2 
	+ \alpha_M\beta_M^*u_M^2 + \alpha_M^*\beta_M u_M^{*2} \Big{]} .
\end{equation}
The dependence on the momentum in functions $u_M$ comes as $k/\mathcal{H}$.
Since $\mu\gg\mathcal{H}$ the argument of this function is always very big, and we can expand these mode functions asymptotically (see Appendix A),
\begin{align}
|u_M|^2 ={}& \frac{1}{2k} \left[ 1 + \mathcal{O}(\mathcal{H}^2) \right],
\\
u_M^2 ={}& \frac{1}{2k} e^{-\frac{2ik}{\mathcal{H}}} 
	\left[ 1 + \mathcal{O}(\mathcal{H}) \right] ,
\end{align}
and we have for the integral
\begin{align}
I_s^{(\mu,\infty)} \approx{}& \frac{1}{2} \int\limits_{\mu}^{\infty} 
	dk \, k^{2s-1}
	\Big{[} 2|\beta_M|^2 + \alpha_M\beta_M^* e^{-\frac{4ik}{\mathcal{H}}}
	+ \alpha_M^*\beta_M e^{\frac{4ik}{\mathcal{H}}} \Big{]}
\nonumber \\
&	\le \frac{1}{2} \int\limits_{\mu}^{\infty} 
	dk \, k^{2s-1} 2|\beta_M|^2 
	+ \left| \frac{1}{2} \int\limits_{\mu}^{\infty} 
	dk \, k^{2s-1} \alpha_M\beta_M^* e^{-\frac{4ik}{\mathcal{H}}} \right|
	+ \left| \frac{1}{2} \int\limits_{\mu}^{\infty} 
	dk \, k^{2s-1} \alpha_M^*\beta_M e^{\frac{4ik}{\mathcal{H}}} \right|
\nonumber \\
&	\le \frac{1}{2} \int\limits_{\mu}^{\infty} 
	dk \, k^{2s-1}
	\Big{[} 2|\beta_M|^2 + 2 |\alpha_M\beta_M^*| \Big{]} ,
\end{align}
The dominant contribution to this integral is a constant (does not evolve in time),
which means that the dominant contribution to energy density and pressure
scales like radiation. Since $\beta$ Bogolyubov coefficient is at least
exponentially suppressed in the UV, $\beta\sim e^{-\tau k}$, 
this contribution has to be small, $\sim \mu^{2s}e^{-\tau \mu}$,
and since it scales like radiation it falls off with time with respect to the background
becoming more and more negligible. Non-negligible contributions, if they exist
come from the IR effects and should be isolated from the IR part of
$\mathcal{I}_s$ integral.

\subsection{IR contribution}

The IR contribution to $\mathcal{I}_s$ integrals,
\begin{equation} \label{Is IR}
\mathcal{I}_s^{(0,\mu)} = \int\limits_{k_0}^{\mu}\! 
	dk\, k^{2s} Z_{Bog.}(k,\eta) ,
\end{equation}
is more difficult to evaluate,
since the IR modes evolve non-adiabatically and are sensitive to the 
expansion history of the Universe (Figure \ref{expansion history}). 
The main problem 
is determining the Bogolyubov coefficients generated by the transitions
between different cosmological eras.

For fast (monotonous) transitions between different periods 
of  constant $\epsilon$ the IR 
modes oscillate slowly enough not to be sensitive to
the details of the transitions. 
Therefore, we expect them to be well described 
by the so-called sudden transition approximation, where the background
$\epsilon$ parameter suddenly jumps from one value to another.
This picture should not be taken too literally as the background model
of the expansion of the Universe,
rather as an expansion in powers of the small time of transition.
Also, this approximation, although being very good in the IR, fails 
in the UV, where it spoils adiabaticity of the UV modes leading to additional
unphysical UV divergences of the energy-momentum tensor (see 
\cite{Glavan:2013mra} for a detailed discussion), and it should
not be used there. Once again,
even though in what follows we treat the background as being
one with sudden transitions, we emphasize
that this is an approximation for the evolution of the IR modes, not an
approximation for the background itself.

\section{Sudden transition approximation}
\label{sec:Sudden transition}

Let us consider a sudden transition from a period of constant $\epsilon_0$
to a period of constant $\epsilon_1$ at some time $\eta=\eta_0$. 
If the scalar mode function before the transition was $U_0(k,\eta)$,
then the  mode function after the transition will be
\begin{equation}
U_1(k,\eta)= \alpha_{1,0}(k) u_1(k,\eta) + \beta_{1,0}(k) u_1^*(k,\eta) ,
\end{equation}
where $u_1$ is the B-D mode function for a constant $\epsilon=\epsilon_1$ 
of the later period, and
$\alpha_{1,0}(k)$, $\beta_{1,0}(k)$ are the Bogolyubov coefficients determined by
the continuity conditions on the mode function and its derivative at the
transition time,
\begin{align}
\alpha_{1,0}(k) ={}& i \Big{[} U_0'(k,\eta_0) u_1^*(k,\eta_0)
	- U_0(k,\eta_0)u_1'^*(k,\eta_0) \Big{]} , \\
\beta_{1,0}(k) ={}& i \Big{[} U_0(k,\eta_0) u_1'(k,\eta_0)
	- U_0'(k,\eta_0)u_1(k,\eta_0) \Big{]} .
\label{alpha1 beta1}
\end{align}

If we are dealing with a sequence of sudden transitions, starting with a B-D state
during the period $\epsilon=\epsilon_0$, it is easy to see that the final Bogolyubov coefficients
during period $\epsilon_n$ will be
\begin{equation} \label{Bogolyubov}
\begin{pmatrix} \alpha_{n,0}(k) \\ \beta_{n,0}(k) \end{pmatrix} 
	= \boldsymbol{T}_n (k,\eta_{n-1})
	\boldsymbol{T}_{n-1}(k,\eta_{n-2}) \dots \boldsymbol{T}_2(k,\eta_1)
	\boldsymbol{T}_1(k,\eta_0) \begin{pmatrix} 1 \\ 0 \end{pmatrix} ,
\end{equation}
where the transfer matrices are
\begin{equation}
\boldsymbol{T}_i(k,\eta_{i-1}) =
	\begin{pmatrix} \alpha_{i,i-1}(k,\eta_{i-1})
		& \beta^*_{i,i-1}(k,\eta_{i-1}) \\
		\beta_{i,i-1}(k,\eta_{i-1})
		& \alpha^*_{i,i-1}(k,\eta_{i-1}) \end{pmatrix} ,
\end{equation}
and the partial Bogolyubov coefficients are defined to be
\begin{align}
\alpha_{i,i-1}(k,\eta_{i-1})
 ={}&  i \Big{[} u_i^*(k,\eta_{i-1})u_{i-1}'(k,\eta_{i-1}) 
	- u_i'^*(k,\eta_{i-1})u_{i-1}(k,\eta_{i-1}) \Big{]} , \label{partial a} \\
\beta_{i,i-1}(k,\eta_{i-1}) ={}&  i \Big{[} u_i'(k,\eta_{i-1}) u_{i-1}(k,\eta_{i-1}) 
	- u_i(k,\eta_{i-1}) u'_{i-1}(k,\eta_{i-1}) \Big{]} , \label{partial b} 
\end{align}
and are composed only of B-D mode functions and their derivatives 
evaluated at the transition time.

\subsection{Bogolyubov coefficients in the UV}

These Bogolyubov coefficients (\ref{Bogolyubov}) should capture the time evolution of the IR modes 
(the definition of which somewhat changes during evolution). But the UV modes 
undergo a different type of evolution, namely adiabatic. Since it costs a lot of
energy to excite the UV modes, we expect them not to have a significant contribution
to the energy density and pressure in the end. Therefore, the exact structure of the
Bogolyubov coefficients in the UV should not be relevant, as long as the unitarity
condition (\ref{unitarity}) is satisfied, and
$|\alpha| \rightarrow 1$, $\beta \rightarrow 0$ faster than any power of
$1/k$ as $k\rightarrow \infty$. A simple way to model this is to construct the full 
Bogolyubov coefficients out of the sudden transition ones in the following manner.
\begin{align}
\beta_{i,i-1} \rightarrow \beta_{i,i-1}\, e^{-k\tau_{i-1}}, \qquad
	  \alpha_{i,i-1} \rightarrow \sqrt{1+\beta_{i,i-1}\, e^{-k\tau_{i-1}}} \,
		\frac{\alpha_{i,i-1}}{|\alpha_{i,i-1}|} ,
\end{align}
They quickly reduce to the sudden transition ones in the IR, 
and are non-adiabatically suppressed
in the UV, and satisfy (\ref{unitarity}) by construction. 
The point of this regularization is not 
to represent exactly the evolution of the mode function, 
but rather it allows us to extract the
leading order behaviour of the energy-momentum tensor 
while making intermediate steps of the
calculation well defined. Parameters $\tau_i$ have clear physical
interpretation as the durations of transitions, and are (in principle)
an independent scale of the system. By introducing them we have
control over the approximation for fast transition, and the 
expansion in $\tau_i$'s
yields the leading behaviour\footnote{In \cite{Aoki:2014ita} the 
sudden transition Bogolyubov coefficients were regulated by cutting them off
sharply at the Hubble rate $\mathcal{H}$ of the transition, 
$\alpha(k)=1,\beta(k)=0$ for $k>\mathcal{H}$. This way the durations of
transitions are no longer independent scales, but are intimately tied with
the Hubble rate of the transition, and they are not short compared 
to the Hubble rate either. Such approach accounts for 
all the terms in the energy density and pressure when it comes to 
their scaling in time (see Appendix D in \cite{Glavan:2013mra}),
but their coefficients (in principle) cannot be calculated reliably because
the two scales describing the transition were assumed the same from the start,
and there is no control over the duration of transition expansion. Though
the leading order results in \cite{Glavan:2013mra} and \cite{Aoki:2014ita} agree}. 
Calculating corrections of order $\tau_i$ and higher to the result does not make
much sense.

\vskip+3cm

\subsection{Bogolyubov coefficients in the deep IR}
\label{sec:Sudden transition IR}

In this subsection we present the IR expansion of the Bogolyubov coefficients
generated by $n$ transitions between periods of constant $\epsilon$.
This was already derived, in a much more general setting of smooth transitions
in \cite{Tsamis:2002qk}. Here we present the expansion for the special
case of sudden transitions in a form suited for our needs.

Consider a sudden transition between two decelerating periods $\epsilon_0$
and $\epsilon_1$. It is convenient to write the B-D
mode function (\ref{u deceleration})
as a power series around $k=0$ (which is just the definition of the Bessel function),
\begin{align}
u(k,\eta) ={}& \sqrt{\frac{\pi}{4(\epsilon\!-\!1)\mathcal{H}}} \frac{i}{\sin(\pi\nu)}
	\left[ J_{-\nu}\left( \frac{k}{(\epsilon\!-\!1)\mathcal{H}} \right)
	- e^{i\pi\nu} J_{\nu}\left( \frac{k}{(\epsilon\!-\!1)\mathcal{H}} \right) \right]
\label{mode functions expanded} \\
&\hskip -1cm =
{} \sqrt{\frac{\pi}{4(\epsilon\!-\!1)\mathcal{H}}} \frac{i}{\sin(\pi\nu)}
	\Bigg{[} \frac{[2(\epsilon\!-\!1)\mathcal{H}]^\nu}{k^\nu}
		S_{-\nu}\left( \frac{k}{(\epsilon\!-\!1)\mathcal{H}} \right)
	- e^{i\pi\nu} \frac{k^\nu}{[2(\epsilon\!-\!1)\mathcal{H}]^\nu}S_{\nu}\left( \frac{k}{(\epsilon\!-\!1)\mathcal{H}} \right) \Bigg{]} ,
\nonumber
\end{align}
and similarly for the case of a $\epsilon<1$ B-D mode function, where
\begin{equation}
S_{\pm\nu}(z) = \sum_{l=0}^{\infty} \frac{(-1)^l}{l! \, \Gamma(l\pm\nu+1)}
	\left( \frac{z}{2} \right)^{2l} =  \frac{1}{\Gamma(1\pm\nu)} +{\cal O}(z^2).
\end{equation}
We will be interested in the leading order term in the Bogolyubov coefficients,
so for non-integer $\nu$ it suffices to keep only the first term in the series 
for $S_{\pm\nu}(z)\simeq 1/\Gamma(1\pm\nu)$
from the start. Therefore, it is enough to work with
\begin{equation}
u_n(k,\eta) = i M_n(\mathcal{H}) k^{-\nu_n}
	- i e^{i\pi\nu_n} P_n(\mathcal{H}) k^{\nu_n}
\end{equation}
for $\epsilon_n>1$, and with
\begin{equation}
u_n(k,\eta) = - i M_n(\mathcal{H}) k^{-\nu_n}
	+ i e^{-i\pi\nu_n} P_n(\mathcal{H}_n) k^{\nu_n}
\end{equation}
for $\epsilon_n<1$, where
\begin{align}
M_n(\mathcal{H}) ={}& \sqrt{\frac{\pi}{2}}
\frac{[2|\epsilon_n\!-\!1| \mathcal{H}]^{\nu_n-\frac12}}{\sin(\pi\nu_n)\Gamma(1\!-\!\nu_n)}
=\frac{\Gamma(\nu_n)}{\sqrt{2\pi}}[2|\epsilon_n\!-\!1|\,\mathcal{H}]^{\nu_n-\frac12}
  ,
\label{Mn}\\
P_n(\mathcal{H}) ={}& \sqrt{\frac{\pi}{2}}
\frac{[2|\epsilon_n\!-\!1| \mathcal{H}]^{-\nu_n-\frac12}}{\sin(\pi\nu_n)\Gamma(1\!+\!\nu_n)} 
=-\frac{\Gamma(-\nu_n)}{\sqrt{2\pi}}[2|\epsilon_n\!-\!1|\,\mathcal{H}]^{-\nu_n-\frac12}
\label{Pn}\,.
\end{align}

Next we show that the structure of the Bogolyubov coefficients does not
change depending on how many sudden transitions there are between
the first and the last period of constant $\epsilon$.
For a sudden transition between two decelerating periods,
from $\epsilon_0$ to $\epsilon_1$, we
can calculate the partial Bogolyubov coefficients defined in (\ref{partial a})
and (\ref{partial b}),
\begin{align}
\alpha_{1,0}(k) ={}& i A_{1,0} k^{-\nu_0-\nu_1}
	- i e^{i\pi\nu_0} B_{1,0} k^{\nu_0-\nu_1}
	- i e^{-i\pi\nu_1} C_{1,0} k^{-\nu_0+\nu_1}
	+i e^{i\pi(\nu_0-\nu_1)} D_{1,0} k^{\nu_0+\nu_1} , 
\label{alpha10}\\
\beta_{1,0}(k) ={}& i A_{1,0} k^{-\nu_0-\nu_1}
	-  ie^{i\pi\nu_0} B_{1,0} k^{\nu_0-\nu_1}
	- i e^{i\pi\nu_1} C_{1,0} k^{-\nu_0+\nu_1}
	+i e^{i\pi(\nu_0+\nu_1)} D_{1,0} k^{\nu_0+\nu_1} ,
\label{beta10}
\end{align}
where
\begin{align}
A_{1,0} ={}& M_0'(\mathcal{H}_0) M_1(\mathcal{H}_0)
	- M_0(\mathcal{H}_0) M_1'(\mathcal{H}_0) , 
\label{A10}\\
B_{1,0} ={}&  P_0'(\mathcal{H}_0) M_1(\mathcal{H}_0)
	- P_0(\mathcal{H}_0) M_1'(\mathcal{H}_0) , 
\label{B10}\\
C_{1,0} ={}& M_0'(\mathcal{H}_0) P_1(\mathcal{H}_0)
	- M_0(\mathcal{H}_0) P_1'(\mathcal{H}_0) , 
\label{C10}\\
D_{1,0} ={}& P_0'(\mathcal{H}_0) P_1(\mathcal{H}_0)
	- P_0(\mathcal{H}_0) P_1'(\mathcal{H}_0) .
\label{D10}
\end{align}
If the state is a B-D one during the $\epsilon_0$-period, the full Bogolyubov
coefficients correspond to the partial ones for one transition
given in Eqs.~(\ref{alpha10}--\ref{beta10}).

For two successive transitions between decelerating periods we have from
(\ref{Bogolyubov})
\begin{align}
\alpha_{2,0}(k) ={}& \alpha_{2,1}(k)\alpha_{1,0}(k)
	+ \beta^*_{2,1}(k)\beta_{1,0}(k)
\nonumber \\
	={}& i A_{2,0} k^{-\nu_0-\nu_2}
	- i e^{i\pi\nu_0} B_{2,0} k^{\nu_0-\nu_2}
	- i e^{-i\pi\nu_2} C_{2,0} k^{-\nu_0+\nu_2}
	+i e^{i\pi(\nu_0-\nu_2)} D_{2,0} k^{\nu_0+\nu_2} , \label{alpha2} \\
\beta_{2,0}(k) ={}& \beta_{2,1}(k) \alpha_{1,0}(k)
	+ \alpha^*_{2,1}(k) \beta_{1,0}(k)
\nonumber \\
	={}& i A_{2,0} k^{-\nu_0-\nu_2}
	- i e^{i\pi\nu_0} B_{2,0} k^{\nu_0-\nu_2}
	- i e^{i\pi\nu_2} C_{2,0} k^{-\nu_0+\nu_2}
	+i e^{i\pi(\nu_0+\nu_2)} D_{2,0} k^{\nu_0+\nu_2} , \label{beta2}
\end{align}
where
\begin{align}
A_{2,0} ={}& 2\sin(\pi\nu_1)\Big{[}B_{2,1}A_{1,0}- A_{2,1}C_{1,0} \Big{]} , 
\label{A20}\\
B_{2,0} ={}& 2\sin(\pi\nu_1)\Big{[}B_{2,1} B_{1,0}-A_{2,1} D_{1,0} \Big{]} , 
\label{B20}\\
C_{2,0} ={}& 2\sin(\pi\nu_1)\Big{[}D_{2,1} A_{1,0}-C_{2,1}C_{1,0}  \Big{]} , 
\label{C20}\\
D_{2,0} ={}& 2\sin(\pi\nu_1)\Big{[} D_{2,1}B_{1,0}-C_{2,1} D_{1,0} \Big{]} 
\,,
\label{D20}
\end{align}
or, when written in a matrix form, 
\begin{equation}
\begin{pmatrix}
   B_{2,0} &  A_{2,0} \\
   D_{2,0} & C_{2,0} \\
\end{pmatrix}
    =  2\sin(\pi\nu_1)
\begin{pmatrix}
   B_{2,1} & A_{2,1} \\
   D_{2,1} & C_{2,1} \\
\end{pmatrix}
\cdot 
\begin{pmatrix}
   1 & 0 \\
   0 & -1 \\
\end{pmatrix}
\cdot
 \begin{pmatrix}
   B_{1,0} & A_{1,0} \\
   D_{1,0} & C_{1,0} \\
\end{pmatrix}
\,.
\label{A-D20:matrix}
\end{equation}
Upon comparing the structure of Eqs.~(\ref{alpha10}--\ref{beta10}) 
with~(\ref{alpha2}--\ref{beta2}) we see that sudden matchings do
not change the structure of the powers of the momentum in the deep IR.
If there are $n$ sudden transitions between decelerating periods,
the leading power of the momentum in the IR is
$k^{-\nu_0-\nu_n}$. The same conclusion stands if some of the
intermediary periods are accelerating ones, it is just that the coefficient
of the leading term changes, but it is straightforward to calculate it.
In the case when one considers many transitions it is convenient to use 
the suitably generalized matrix form~(\ref{A-D20:matrix}), which for 
$n$ matchings becomes, 
\begin{equation}
\begin{pmatrix}
   B_{n,0} &  A_{n,0} \\
   D_{n,0} & C_{n,0} \\
\end{pmatrix}
    =  
   \left[ \prod_{j=1}^{n-1}2\sin(\pi\nu_j)
  \begin{pmatrix}
   B_{j+1,j} & -A_{j+1,j} \\
   D_{j+1,j} & -C_{j+1,j} \\
\end{pmatrix}
\right]
\cdot
 \begin{pmatrix}
   B_{1,0} & A_{1,0} \\
   D_{1,0} & C_{1,0} \\
\end{pmatrix}
\,.
\label{A-D20:matrix:n}
\end{equation}

These coefficients contain many terms with different powers of the Hubble rates
at points of transition. If there exist a clear hierarchy between these Hubble
rates many of the terms can be neglected as subdominant ones, and coefficients
simplify significantly.

\section{Calculating the IR of integrals $\mathcal{I}_s$}
\label{sec:Calculating IR}

In this section we approximate and calculate integrals (\ref{Is IR}).
Since the integration range here is $(k_0,\mu)$, and 
$k_0\ll\mathcal{H},\mathcal{H}_0\ll \mu \ll \mathcal{H}_2\ll \mathcal{H}_1$
(see Figure \ref{Hubble rate}),
the ratios $k/\mathcal{H}_1$ and $k/\mathcal{H}_2$ are very small,
and we expand parts of the Bogolyubov coefficients in powers of this ratio,
keeping only the leading term. In the end this corresponds to 
expanding the full result
in powers of $(\mathcal{H},\mathcal{H}_0)/(\mathcal{H}_1,\mathcal{H}_2$).

Bogolyubov coefficients can be written in terms of partial ones as as
\begin{align}
\alpha_{3,0} ={}& \alpha_{3,1} \alpha_{1,0} + \beta_{3,1}^* \beta_{1,0} , \\
\beta_{3,0} ={}& \alpha_{3,1}^*\beta_{1,0} + \beta_{3,1}\alpha_{1,0} .
\end{align}
The dependence on $k$ in $\alpha_{3,1}$ and $\beta_{3,1}$ appears only as
$k/\mathcal{H}_1$ and $k/\mathcal{H}_2$, which means that we can expand these,
keeping only the leading order contribution, which is, as derived in Section
V.B,
\begin{equation}
\alpha_{3,1} = \beta_{3,1} = \frac{i A_{3,1}}{k^{\nu_I+\nu_M}} = - \beta_{3,1}^* ,
\end{equation}
where $A_{3,1}$ is defined as in (\ref{A-D20:matrix:n}). In this particular case it is
\begin{equation}
A_{3,1} = - \frac{1}{\pi}\,
	 2^{\nu_I-\frac{5}{2}} \, \Gamma(\nu_I) \Gamma(\nu_M)
	(1-\epsilon_I)^{\nu_I+\frac{1}{2}} \left( \nu_I - \frac{1}{2} \right)
	\left( \nu_M+\frac{3}{2} \right) \mathcal{H}_1^{\nu_I+\frac{1}{2}}
	\mathcal{H}_2^{\nu_M-\frac{1}{2}}
\end{equation}
to leading order in the ratio $\mathcal{H}_2/\mathcal{H}_1\ll1$.
This allows us to write the full coefficients as
\begin{equation}
\alpha_{3,0} = \beta_{3,0} = \beta_{3,1} (\alpha_{1,0} - \beta_{1,0}) .
\end{equation}
The fact that, to leading order, $\alpha_{3,0}=\beta_{3,0}$ simplifies the 
form of the integrand function,
\begin{equation}\label{Z approx1}
Z_{Bog.}(k,\eta) =4 |\beta_{3,0}(k)|^2\, \Re^2[u_M(k,\eta)] 
	= \frac{4 |A_{3,1}|^2}{k^{2\nu_I+2\nu_M}} \,
	|\alpha_{1,0}-\beta_{1,0}|^2\, \Re^2[u_M(k,\eta)] .
\end{equation}
This expression further simplifies if we write out $\alpha_{1,0}$ and $\beta_{1,0}$ in
terms of the B-D mode functions, using the fact that $u_R'(k,\eta)=-ik\, u_R(k,\eta)$,
\begin{equation}
|\alpha_{1,0}(k)-\beta_{1,0}(k)|^2 = \frac{2}{k} \left[ \Re^2[u_I'(k,\eta_0)] 
	+ k^2 \Re^2[u_I(k,\eta_0)] \right] ,
\end{equation}
which can be rewritten using the equation of motion as
\begin{equation} \label{84}
|\alpha_{1,0}(k)-\beta_{1,0}(k)|^2
	= \frac{2}{k} \left[ \frac{1}{2} \frac{\partial^2}{\partial\eta_0^2} 
		+ 2k^2 + f_I(\eta_0) \right] \Re^2[u_I(k,\eta_0)] ,
\end{equation}
where
\begin{equation}
f_I(\eta_0) = -(1-6\xi)(2-\epsilon_I)\mathcal{H}_0^2 .
\end{equation}
This gives us a very convenient representation for the leading order term of the
integrand 
\begin{equation} \label{Z approx}
Z_{Bog.}(k,\eta) = \frac{8 |A_{3,1}|^2}{k^{1+2\nu_I+2\nu_M}}
	\left[ \frac{1}{2}\frac{\partial^2}{\partial\eta_0^2} 
		+ 2k^2 + f_I(\eta_0) \right]
	 \Re^2[u_I(k,\eta_0)] \Re^2[u_M(k,\eta)] ,
\end{equation}
The integral over the integrand above is IR finite
and we set $k_0\rightarrow0$ right away (the terms
that cancel the IR divergences from the B-D contribution to energy density
and pressure come from the subleading contributions in the 
expansion in ratios of Hubble rates).

The derivatives with respect to $\eta_0$ can be taken out  of the integral,
and we have for the $\mathcal{I}_s$ integrals
\begin{equation} \label{IinJ}
\mathcal{I}_s^{(0,\mu)} = 8|A_{3,1}|^2 
	\left\{ 2 \mathcal{J}_{s+1}
	+ \left[ \frac{1}{2}\frac{\partial^2}{\partial\eta_0^2}+f_I(\eta_0) \right]
	\mathcal{J}_s \right\} ,
\end{equation}
where the $\mathcal{J}_s$ integrals are defined to be
\begin{align} \label{J integral}
\mathcal{J}_s ={}& 
	\int\limits_{0}^{\mu}\! dk\, k^{2s-1-2\nu_I-2\nu_M} \,
	\Re^2[u_I(k,\eta_0)] \, \Re^2[u_M(k,\eta)]
\nonumber \\
={}& \frac{\pi^2}{8(1-\epsilon_I)\mathcal{H}_0\mathcal{H}}
	\int\limits_{0}^{\mu}\! dk\, k^{2s-1-2\nu_I-2\nu_M} \,
	J_{\nu_I}^2\left( \frac{k}{(1-\epsilon_I)\mathcal{H}_0} \right)
	J_{\nu_M}^2 \left( \frac{2k}{\mathcal{H}} \right)  , \qquad s=1,2,3 .
\end{align}
As they stand, $\mathcal{J}_s$ integrals cannot be solved analytically,
because of the finite upper limit of integration $\mu$. But we are not 
interested in the dependence on $\mu$, since this is just a fiducial scale
we have introduced to facilitate the approximation of the full integral.
What we are interested in is extracting the $\mu$-independent contribution,
which correspond to expanding the integrals above in powers of
$1/\mu$.\footnote{Formally, one can evaluate $\mathcal{J}_s$ integrals by
writing them as $\int_0^\mu=\int_0^\infty - \int_\mu^\infty$.
The former integrals contains the leading ($\mu$-independent )
contribution, while the latter integral yields an asymptotic series in
powers of $1/\mu$. The hope is that this series can be (Borrel)
resummed, yielding an expression that gets canceled by the $\mu$-dependent
UV contribution.} 
The leading order term is $(1/\mu)^0$, and the $\mu$-dependent
terms should cancel against the terms coming from the UV, order by order. 
In practice this
cancellation is hard to prove exactly since different approximations
are used to calculate the UV and the IR, and all orders should be calculated 
explicitly. 

The leading order contribution to integrals (\ref{J integral}) corresponds to 
extending the limit of integration $\mu\rightarrow\infty$. Again, 
this does not correspond to capturing the UV contribution correctly
(which we argued is irrelevant),
but rather to a leading approximation for the IR contribution.
These integrals exist, and for $\mathcal{H}>2(1-\epsilon_I)\mathcal{H}_0$
they are
\begin{align} \label{J early}
\mathcal{J}_s \approx{}&
	\frac{2^{-3-2s+2\nu_I+2\nu_M} \sqrt{\pi} \ 
		\Gamma( s-\frac{1}{2}-\nu_I )
		\Gamma(1-s+\nu_I+\nu_M)}
	{\Gamma( \frac{3}{2}-s+\nu_I+\nu_M )
		\Gamma( \frac{3}{2}-s+\nu_I+2\nu_M )}
	\, \mathcal{H}^{-2+2s-2\nu_I-2\nu_M}
\nonumber \\
& \times {}_4F_3\Bigg{(} \frac{1}{2}, \frac{1}{2}-\nu_I, \frac{1}{2}+\nu_I,
	1-s+\nu_I+\nu_M; \frac{3}{2}-s+\nu_I, \frac{3}{2}-s+\nu_I+\nu_M,
\nonumber \\
&	\qquad \qquad	\frac{3}{2}-s+\nu_I+2\nu_M; 
	\frac{4(1-\epsilon_I)^2\mathcal{H}_0^2}{\mathcal{H}^2} \Bigg{)}
\nonumber \\
& + \frac{\pi^{3/2}\, \Gamma(s) \Gamma( \frac{1}{2}-s+\nu_I )}
	{16\, \Gamma^2(1+\nu_M) \Gamma(1-s+\nu_I) \Gamma(1-s+2\nu_I)}
	[(1-\epsilon_I)\mathcal{H}_0]^{-1+2s-2\nu_I} \mathcal{H}^{-1-2\nu_M}
\nonumber \\
& \times {}_4F_3 \Bigg{(} s, s-2\nu_I, s-\nu_I, \frac{1}{2}+\nu_M;
	\frac{1}{2}+s-\nu_I, 1+\nu_M, 1+2\nu_M;
	\frac{4(1-\epsilon_I)^2\mathcal{H}_0^2}{\mathcal{H}^2} \Bigg{)} ,
\end{align}
and for $\mathcal{H}<2(1-\epsilon_I)\mathcal{H}_0$
\begin{align} \label{J late}
\mathcal{J}_s \approx{}& 
	\frac{\sqrt{\pi} \,\Gamma( s-\frac{1}{2}-\nu_M )
		\Gamma( 1-s+\nu_I+\nu_M )}
	{32\,\Gamma( \frac{3}{2}-s+\nu_I+\nu_M )
		\Gamma(\frac{3}{2}-s+2\nu_I+\nu_M)}
	[(1-\epsilon_I)\mathcal{H}_0]^{-2+2s-2\nu_I-2\nu_M}
\nonumber \\
& \times {}_4F_3 \Bigg{(} \frac{1}{2}, \frac{1}{2}-\nu_M, \frac{1}{2}+\nu_M,
	1-s+\nu_I+\nu_M; \frac{3}{2}-s+\nu_M, \frac{3}{2}-s+\nu_I+\nu_M,
\nonumber \\
&	\qquad \qquad  \frac{3}{2}-s+2\nu_I+\nu_M; 
	\frac{\mathcal{H}^2}{4(1-\epsilon_I)^2\mathcal{H}_0^2} \Bigg{)}
\nonumber \\
& + \frac{2^{-4-2s-2\nu_I+2\nu_M} \pi^{3/2} \, \Gamma(s) 
		\Gamma( \frac{1}{2}-s+\nu_M )}
	{\Gamma^2(1+\nu_I) \Gamma(1-s+\nu_M)\Gamma(1-s+2\nu_M)}
	[(1-\epsilon_I)\mathcal{H}_0]^{-1-2\nu_I} \mathcal{H}^{-1+2s-2\nu_M}
\nonumber \\
& \times {}_4F_3 \Bigg{(} s, \frac{1}{2}+\nu_I, s-2\nu_M, s-\nu_M;
	1+\nu_I,1+2\nu_I, \frac{1}{2}+2-\nu_M; 
	\frac{\mathcal{H}^2}{4(1-\epsilon_I)^2\mathcal{H}_0^2} \Bigg{)} ,
\end{align}
where ${}_4F_3$ is the generalized hypergeometric function. 
These two solutions on two different regions actually represent one continuous
solution,
which we were not able to write as one function over the whole interval.
Integrals $\mathcal{I}$ follow from these via expression (\ref{IinJ}).
The derivatives with respect to $\eta_0$ can be replaced by more convenient
ones with respect to $\mathcal{H}_0$. Remember that 
derivatives with respect to $\eta_0$ are actually derivatives 
of $u_I(k,\eta)$ evaluated at $\eta_0$. Since $u_I$ depends on conformal time 
only via $\mathcal{H}$ we can exploit the chain rule,
\begin{equation} \label{time_derivative}
\frac{\partial}{\partial\eta} = \mathcal{H}' \frac{\partial}{\partial\mathcal{H}}
=(1-\epsilon_I)\mathcal{H}^2 \frac{\partial}{\partial\mathcal{H}} \ ,
\end{equation}
and evaluate derivatives at $\eta_0$ to yield
\begin{align} \label{IfromJ}
\mathcal{I}_s \approx{}& \mathcal{I}_s^{(0,\mu)} 
	= 8|A_{3,1}|^2 \Bigg{\{} 2\mathcal{J}_{s+1} + 
	\Bigg{[} \frac{1}{2} (1-\epsilon_I)^2\mathcal{H}_0^4
		\frac{\partial^2}{\partial\mathcal{H}_0^2}
	 + (1-\epsilon_I)^2 \mathcal{H}_0^3
		\frac{\partial}{\partial\mathcal{H}_0}
\nonumber \\
&	\qquad \qquad \qquad \qquad \qquad \qquad
	 - (1-6\xi) (2-\epsilon_I)\mathcal{H}_0^2 \Bigg{]} \mathcal{J}_s \Bigg{\}} .
\end{align}

\section{Energy density and pressure in matter era}
\label{sec:Results}

Energy density and pressure follow from $\mathcal{I}$ integrals
given in (\ref{IfromJ}) via (\ref{rhoQinI}) and (\ref{pQinI}). 
Derivatives with respect to
$\eta$ there can be replaced by more convenient derivatives with respect
to $\mathcal{H}$ (utilizing (\ref{time_derivative}) again),
\begin{align}
\rho_q ={}& \frac{\hbar}{c^3} \times 
	\frac{1}{4\pi^2a^4} \left\{ 2\mathcal{I}_2 +
	\frac{1}{2} \left[ (1-6\xi)\mathcal{H}^2 
	+ \frac{3}{2}(1-4\xi)\mathcal{H}^3\frac{\partial}{\partial\mathcal{H}}
	+ \frac{1}{4}\mathcal{H}^4 \frac{\partial^2}{\partial\mathcal{H}^2} 
	\right] \mathcal{I}_1 \right\} , \label{rho H}
\\
p_q ={}& \frac{\hbar}{c^3} \times 
	\frac{1}{4\pi^2a^4} \left\{ \frac{2}{3}\mathcal{I}_2 +
	\frac{1}{2} \left[ (1-6\xi)\mathcal{H}^2 
	+ \frac{1}{2}(3-16\xi)\mathcal{H}^3\frac{\partial}{\partial\mathcal{H}}
	+ \frac{1}{4}\mathcal{H}^4(1-4\xi) \frac{\partial^2}{\partial\mathcal{H}^2} 
	\right] \mathcal{I}_1 \right\} , \label{p H}
\end{align}
where we have restored the units. By taking the limit of these expressions
$\epsilon_I\rightarrow0$, $\xi\rightarrow0$, 
$\mathcal{H}_0\rightarrow0$ we precisely recover the main result
of \cite{Glavan:2013mra} where the backreaction scales just as the 
background during matter period.
In order to evaluate these expressions in a more general setting presented here,
 we have to specify all the parameters
$\mathcal{H}_0,\mathcal{H}_1,\mathcal{H}_2,\epsilon_I$. These have to 
be connected with the usual cosmological parameters which were taken
from \cite{Ade:2013ktc}.

The value of the Hubble rate today is 
$\mathcal{H}=H=68 \mathrm{\frac{km}{Mps\, s}}$
(for $a=1$ today). The redshift of radiation-matter equality 
$z_{eq}=3270$ fixes the
duration of the matter period, and hence $\mathcal{H}_2$.
The requirement of at least minimal duration of inflationary period,
namely $\mathcal{H}_0=\mathcal{H}$ and the amplitude of the 
scalar perturbations of the CMB fix the duration of inflation and 
radiation period, up to $\epsilon_I$. That is fixed by the
tensor to scalar perturbation ratio in the CMB which was claimed to be
measured recently ($r=0.2$ \cite{Ade:2014xna}) giving $\epsilon_I=0.01$. 
How $\mathcal{H}_i$ 
are determined from these quantities is given in Appendix \ref{parameters}.
The parameters that are left unspecified (whose parameter space we will explore)
are the duration of inflation, expressed in its number of e-foldings $N_I$,
and the non-minimal coupling $\xi$. 

\begin{turnpage}

\begin{figure}[h]

\vskip-2cm
\begin{centering}

\subfloat{\includegraphics[width=4.5cm]{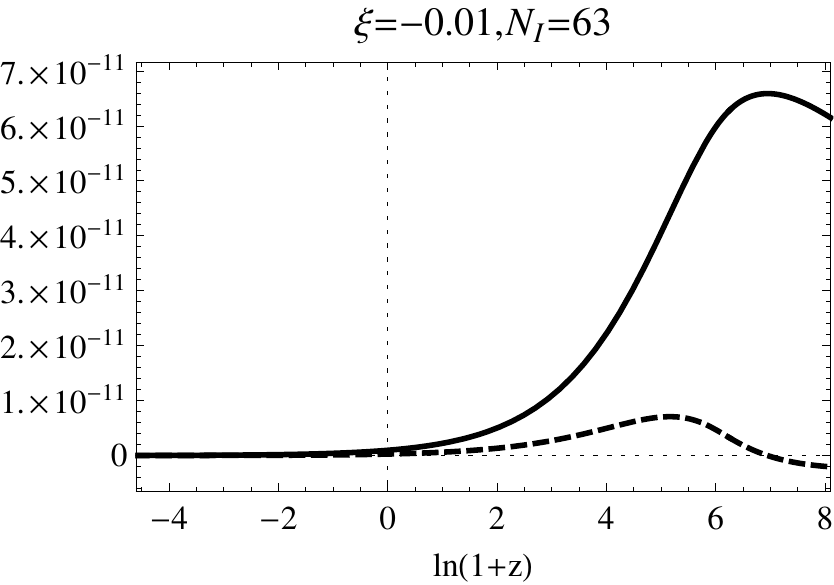}} \hfill
\subfloat{\includegraphics[width=4.5cm]{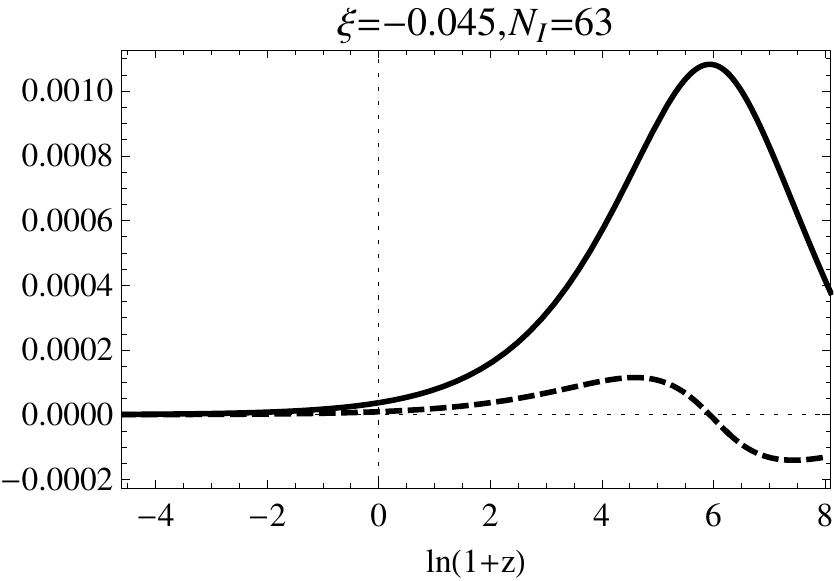}} \hfill
\subfloat{\includegraphics[width=4.5cm]{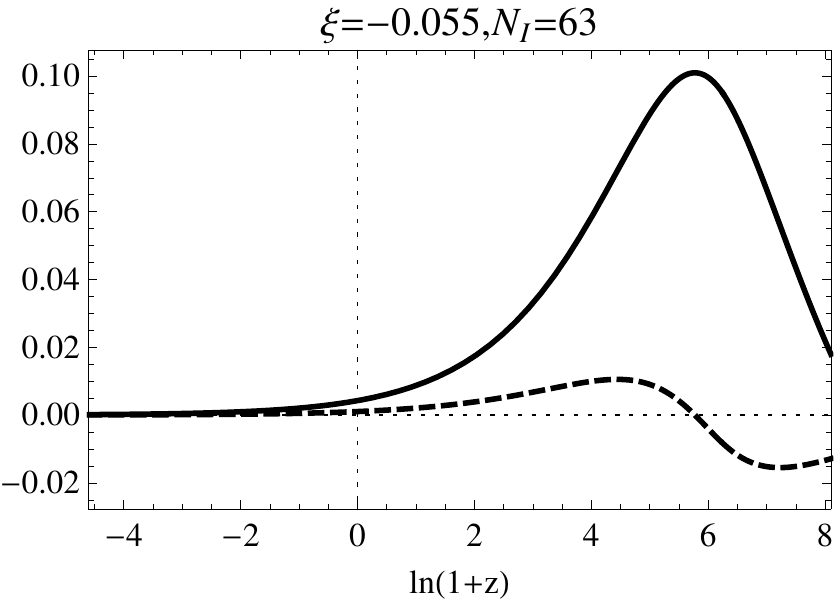}} \hfill
\subfloat{\includegraphics[width=4.5cm]{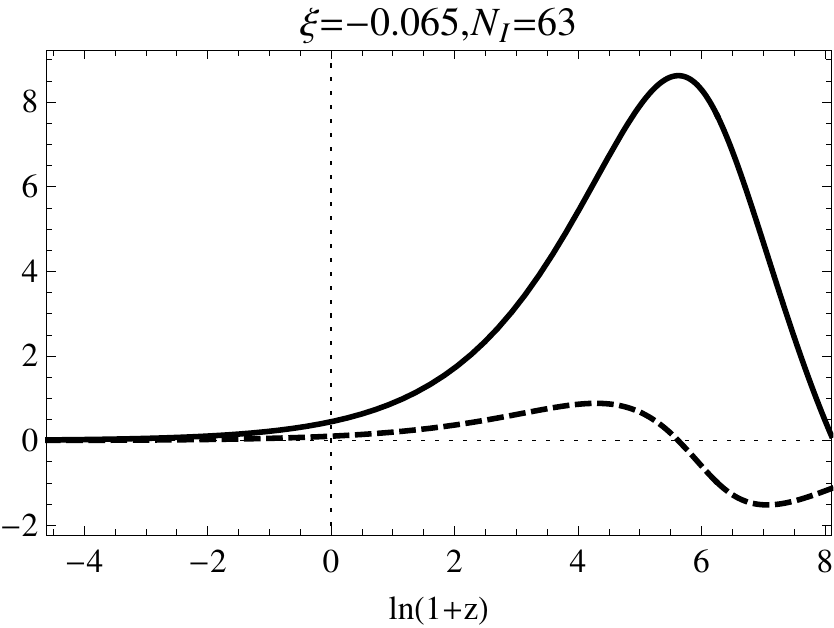}} \hfill
\subfloat{\includegraphics[width=4.5cm]{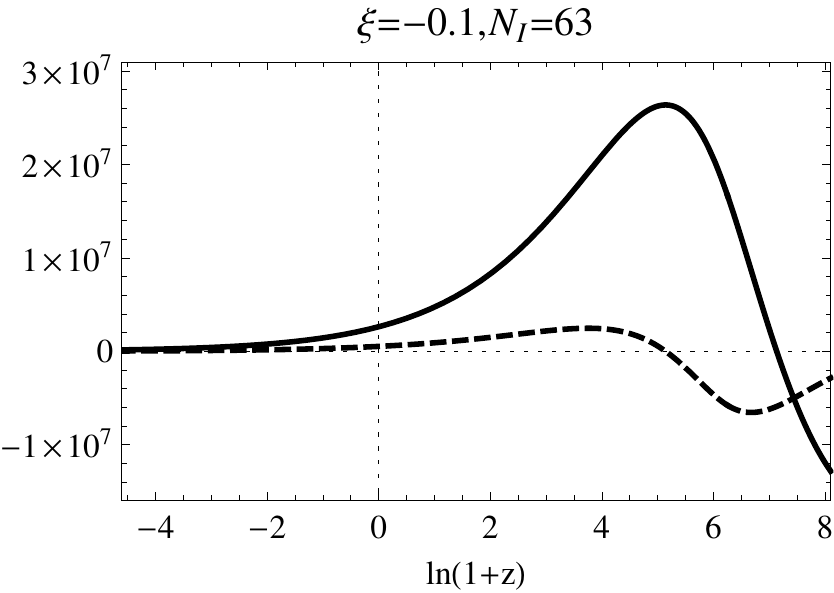}} 
\\
\subfloat{\includegraphics[width=4.5cm]{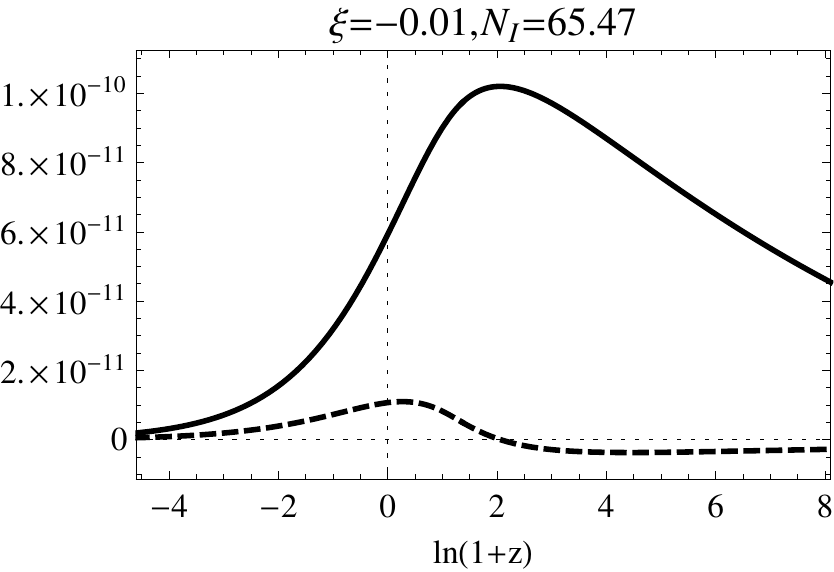}} \hfill
\subfloat{\includegraphics[width=4.5cm]{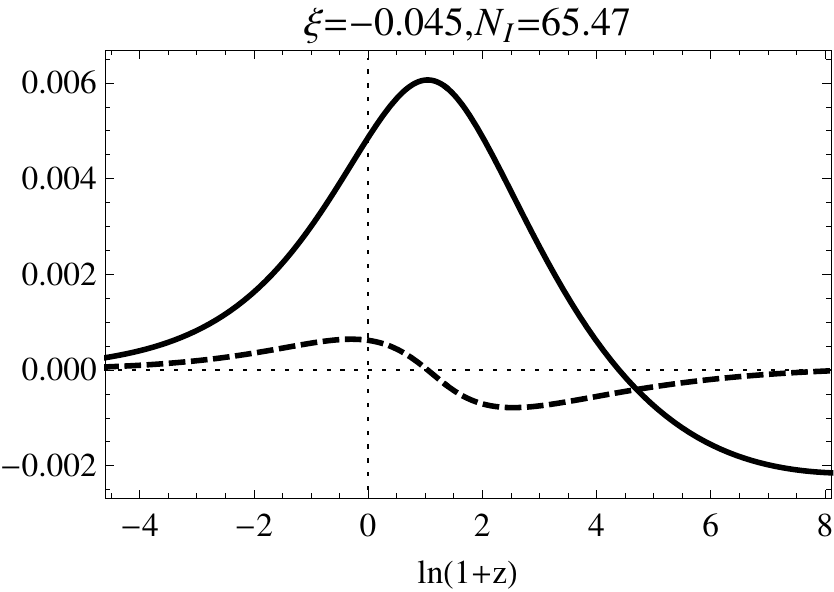}} \hfill
\subfloat{\includegraphics[width=4.5cm]{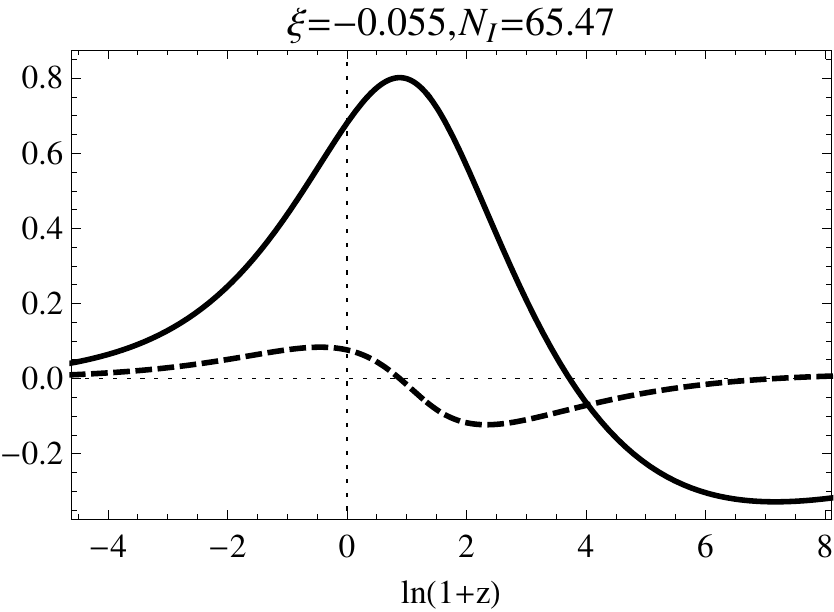}} \hfill
\subfloat{\includegraphics[width=4.5cm]{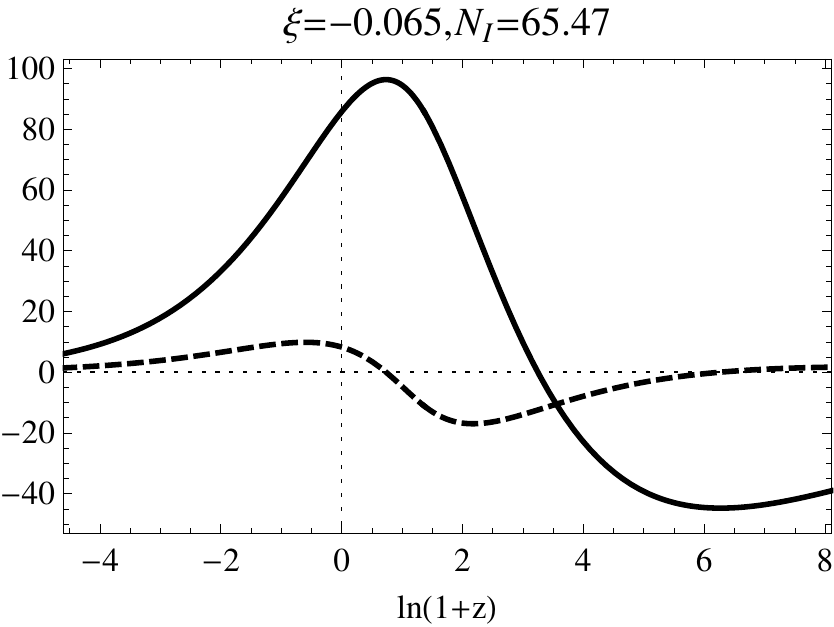}} \hfill
\subfloat{\includegraphics[width=4.5cm]{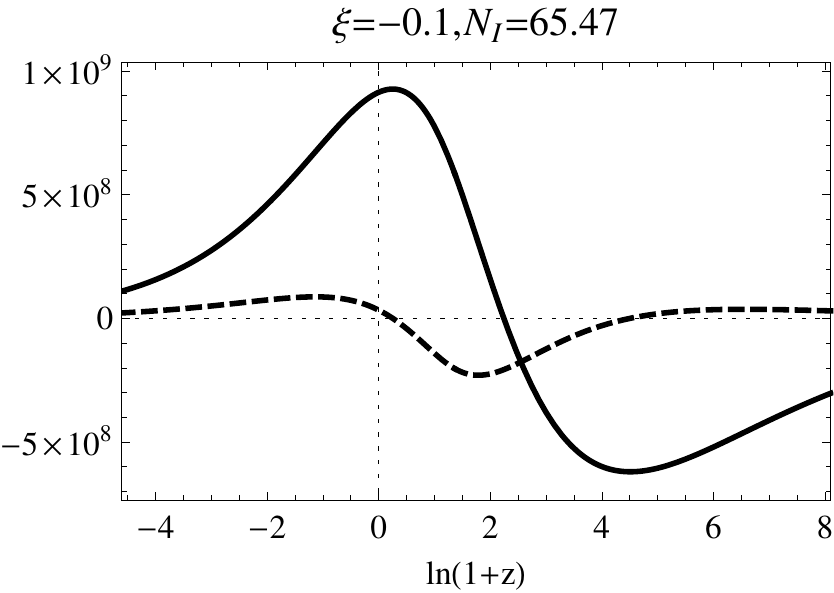}} \hfill
\\
\subfloat{\includegraphics[width=4.5cm]{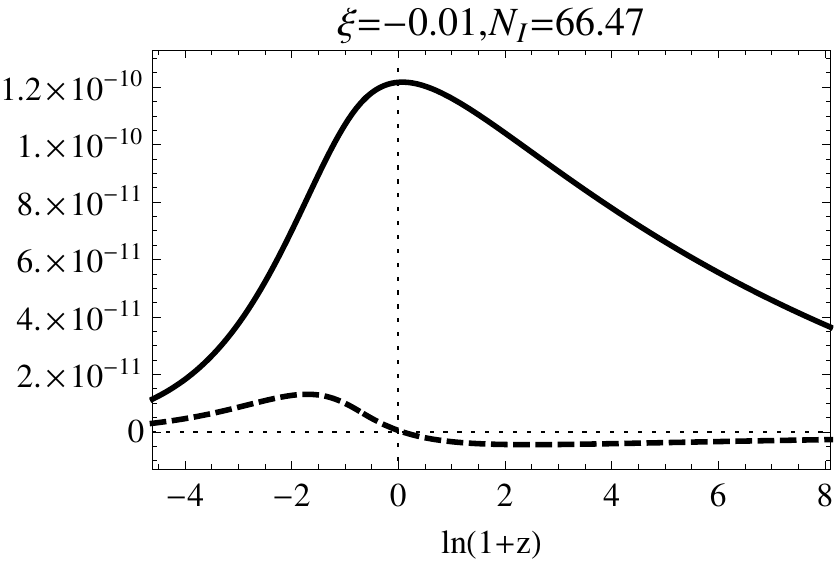}} \hfill
\subfloat{\includegraphics[width=4.5cm]{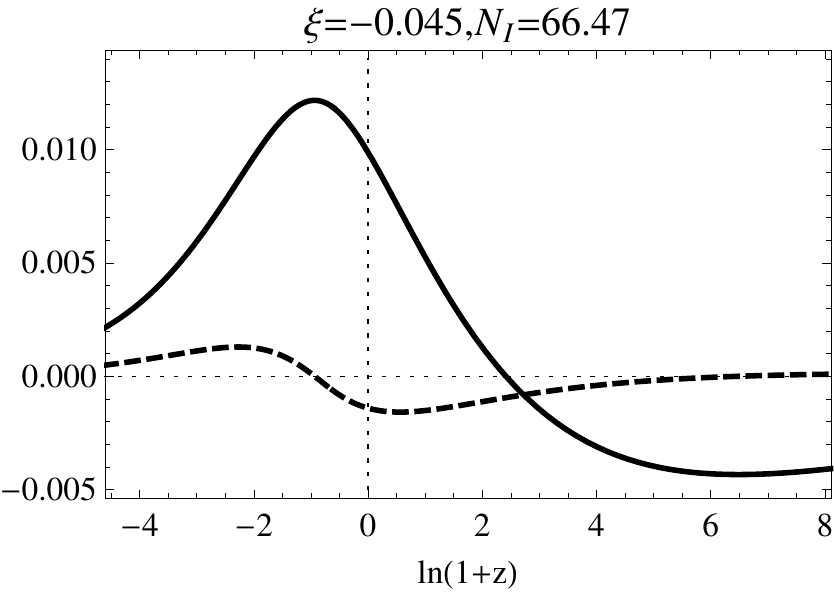}} \hfill
\subfloat{\includegraphics[width=4.5cm]{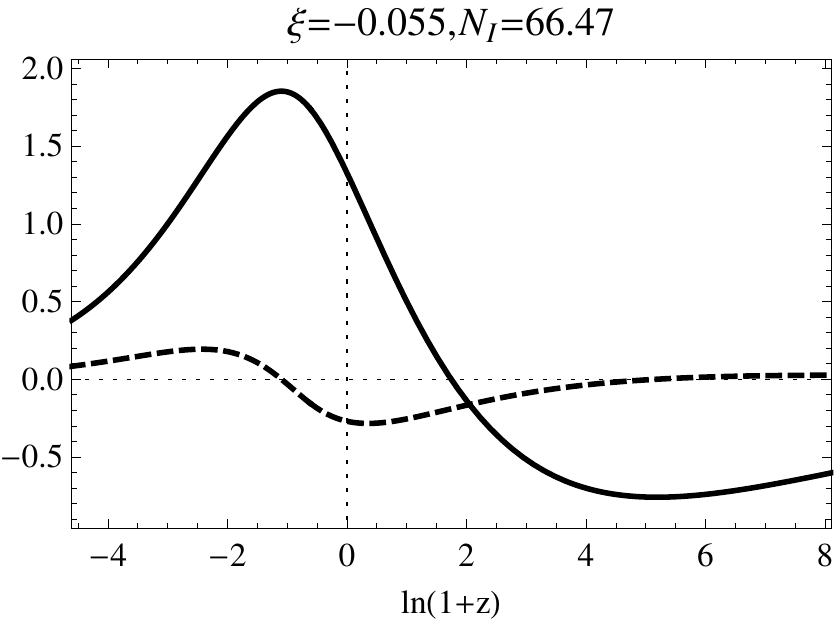}} \hfill
\subfloat{\includegraphics[width=4.5cm]{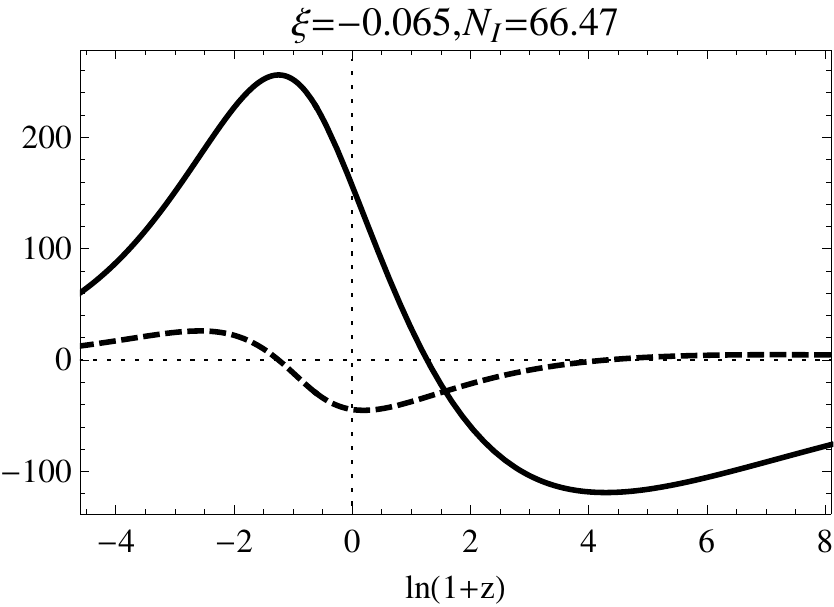}} \hfill
\subfloat{\includegraphics[width=4.5cm]{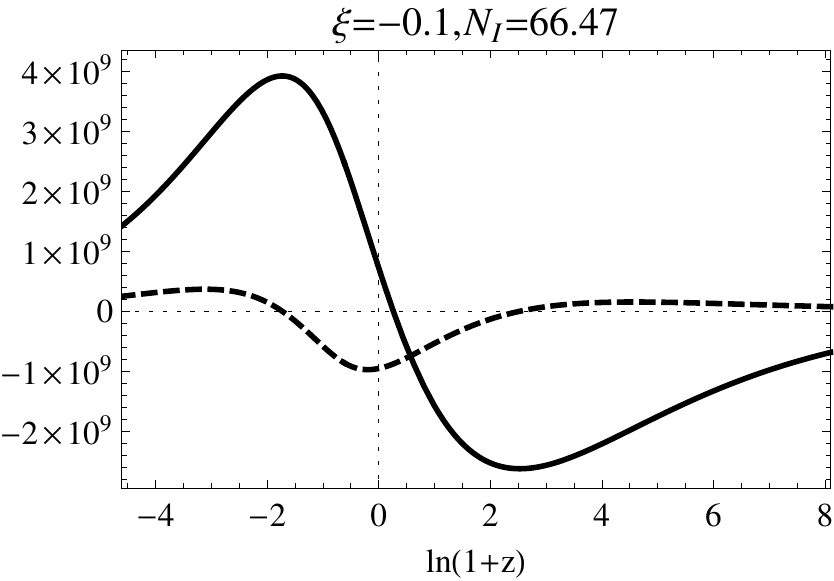}}
\\
\subfloat{\includegraphics[width=4.5cm]{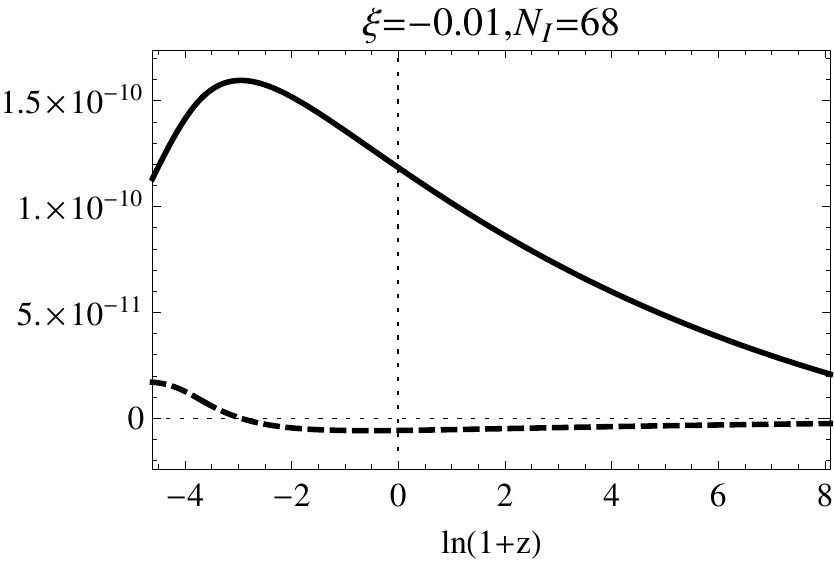}} \hfill
\subfloat{\includegraphics[width=4.5cm]{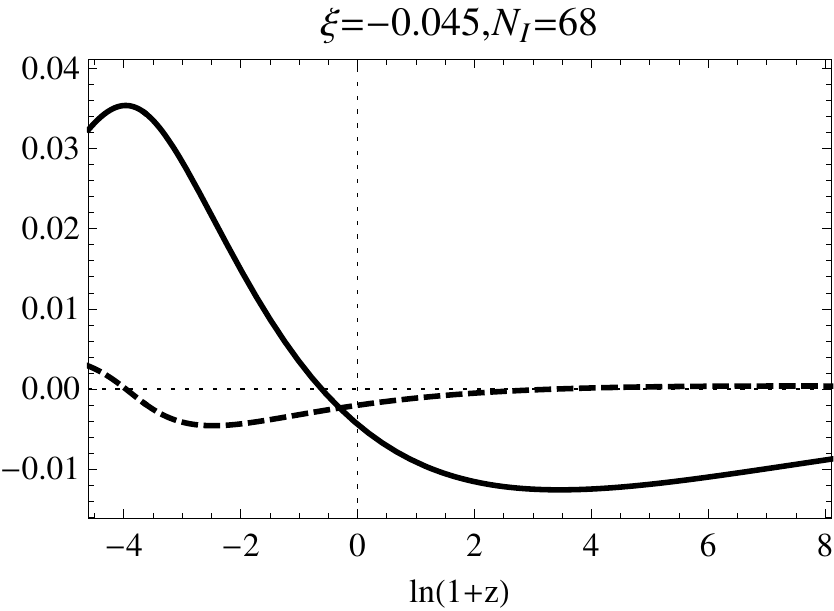}} \hfill
\subfloat{\includegraphics[width=4.5cm]{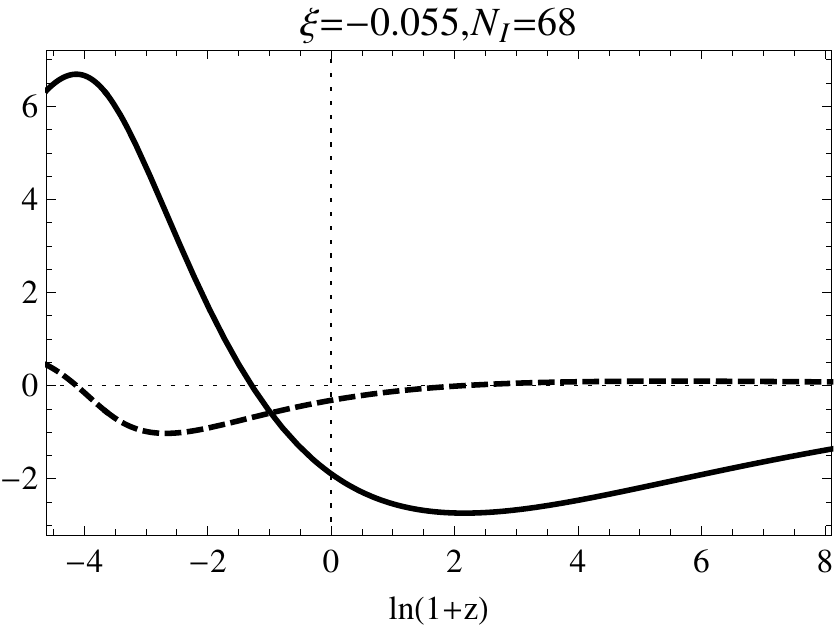}} \hfill
\subfloat{\includegraphics[width=4.5cm]{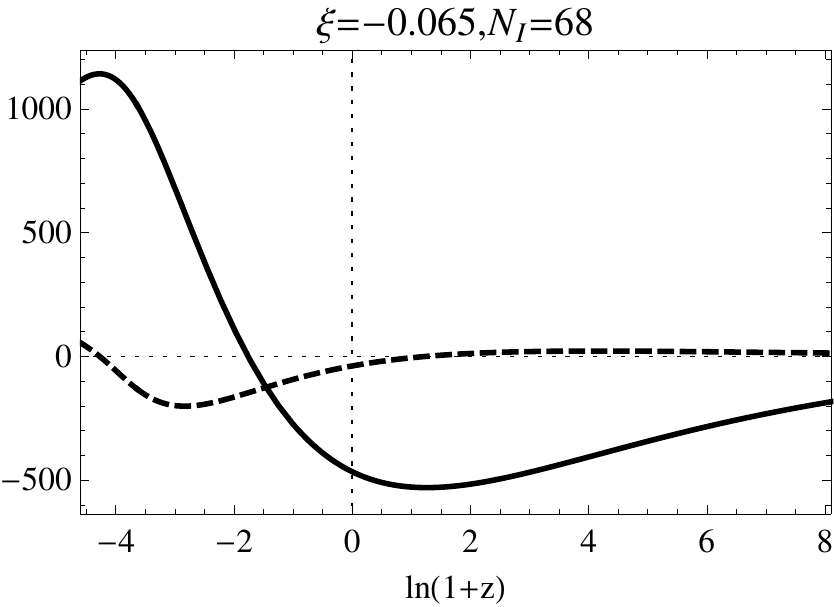}} \hfill
\subfloat{\includegraphics[width=4.5cm]{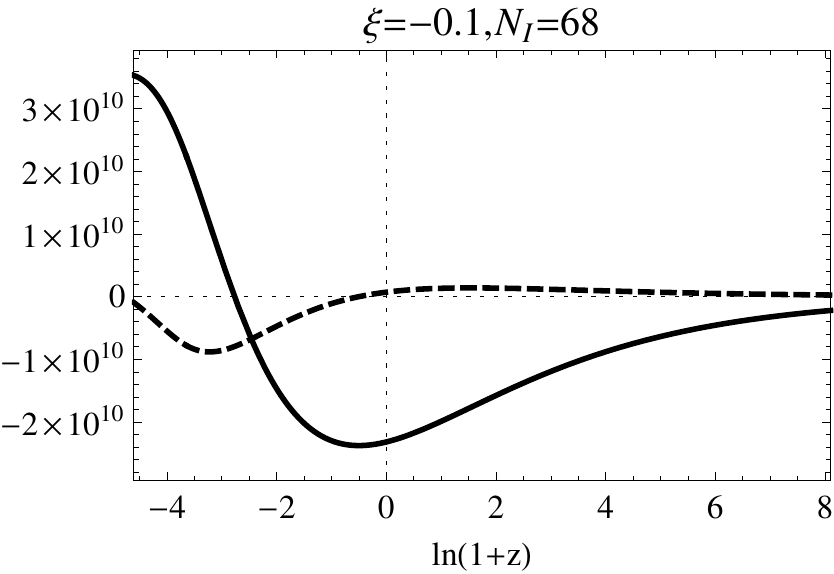}}

\end{centering}

\caption{The backreaction during matter era depending on non-minimal coupling
$\xi$ and number of e-foldings of inflation $N_I$. Solid curves represent the
ratio $\rho_q/\rho_b$, and dashed curves ratio $p_q/\rho_b$ (vertical axis).
The horizontal axis represents the time in terms of the logarithm of redshift.
Redshift $z=0$ corresponds to today, larger values to the past, and negative
values to the future.}

\label{parameter_space_plots}

\end{figure}

\end{turnpage}

Few plots of the backreaction during matter era 
for different parameters ($N_I,\xi$) are presented in Figure 
\ref{parameter_space_plots} to give a feeling for the dependence on parameters.
They all show transient behavior in a form of a pronounced peak 
of the ratio $\rho_q/\rho_b$ around the time when $\mathcal{H}$ becomes
comparable to $\mathcal{H}_0$. The amplitude of the effect is very sensitive 
to changing $\xi$. Changing $N_I$ modifies the amplitude somewhat, but 
also moves the onset of the transient effect, since it changes the time when 
$\mathcal{H}\sim\mathcal{H}_0$.

\begin{figure}[h]
\centering
\includegraphics[width=8cm]{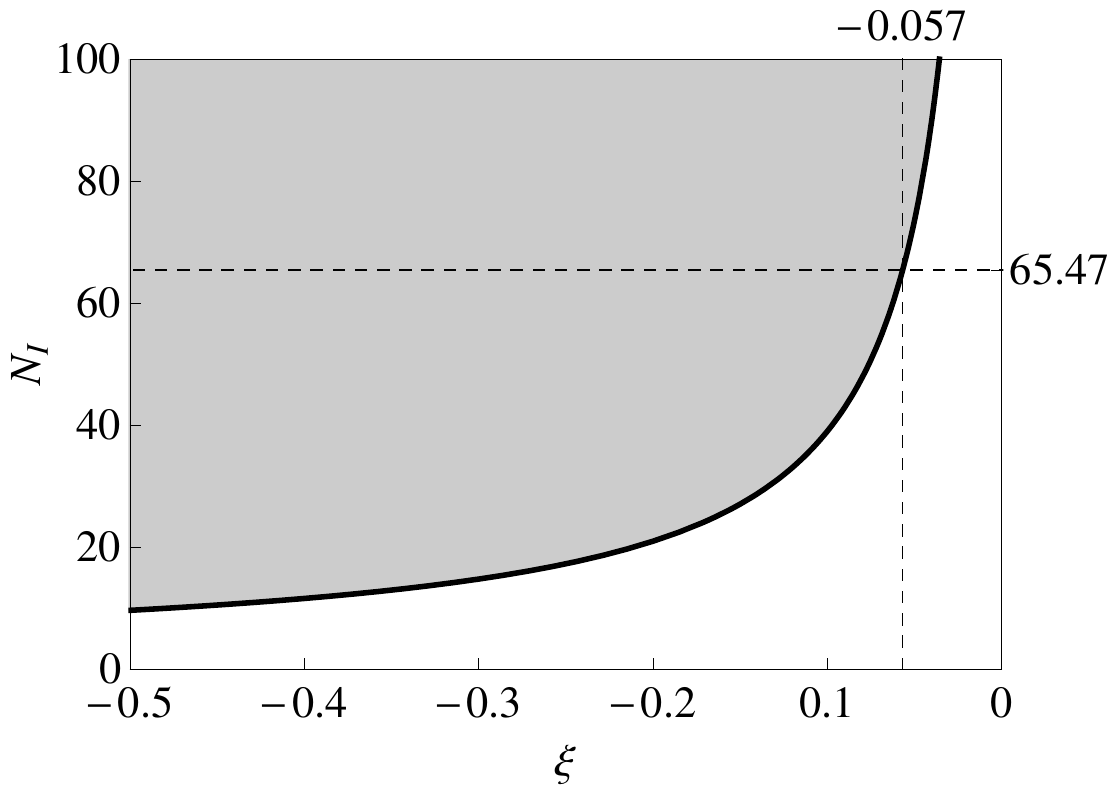}
\caption{The condition $\rho_q/\rho_b=1$ at the end of inflation depending
on the non-minimal coupling and the number of e-foldings of inflation $N_I$.
The shaded region represents the parameter space where this ratio is larger than 1
and where the backreaction influences the evolution during inflation. We do not
study this case in this work. We study the unshaded region, above the dashed 
line of minimal inflation, $N_{I,min}\approx 65.47$ 
(taking the BICEP2 result seriously). This puts the constraint
on non-minimal coupling interesting for this study to be $-0.057<\xi<0$.}
\label{inflation constraint}
\end{figure}

Obviously, there is little point considering parameters for which the backreaction
is always a few orders of magnitude smaller than the 
background. But, also, there is no
point in examining the behavior of backreaction during matter period when it is 
much larger than the background. That signals it must have become
important much before the onset of matter period. That gives us a way to constrain
our parameters. First of all, we require the backreaction to be small during inflation.
The bound for this is $\rho_q/\rho_b\sim 1$ at the end of
inflation\footnote{Studying 
the parameter space for which backreaction becomes large
during inflation would be interesting in its own right. Especially since
in that case it exhibits the behaviour of a perfect fluid with $w<-1$ and
negative contribution to the energy density, which signals it would work
towards slowing down inflation, and perhaps ending it. This possibility was
examined in the case of scalars \cite{Abramo:2001dd},
\cite{Abramo:1998hi}, \cite{Kahya:2009sz}, and in the case of gravitons
 \cite{Tsamis:1996qm}
, \cite{Tsamis:1996qq}. 
Here a full self-consistent (numerical) solution should be found
to be sure how it influences 
the dynamics of the expansion (much like in \cite{Suen:1987gu}).
}.
This ratio can be inferred from expression (59) of reference \cite{Janssen:2009nz},
and is plotted here in Figure \ref{inflation constraint}. 

Even though studying
strong backreaction during inflation is an interesting topic by itself,
here we concentrate on studying backreaction
that becomes strong only in late time matter era. During radiation-dominated 
era the ratio $\rho_q/\rho_b$ does not change
appreciably (that result is presented in Appendix C).
If, for given $\xi$ and $N_I$ backreaction becomes non-negligible during 
matter period, but is still in perturbative range $(\rho_q/\rho_b<1)$, we can
hope to get some intuition about its influence on the background 
evolution if we plot the perturbatively corrected $\epsilon$ parameter,
which we infer from the perturbatively corrected Friedmann equtions,
\begin{equation} \label{eps tot}
\epsilon_{tot} = 1 - \frac{\mathcal{H}'}{\mathcal{H}^2}
	= \frac{3}{2}\left( 1 + \frac{p_q}{\rho_b+\rho_q} \right) .
\end{equation}

\begin{figure}[h]

\subfloat[]{\includegraphics[width=8cm]{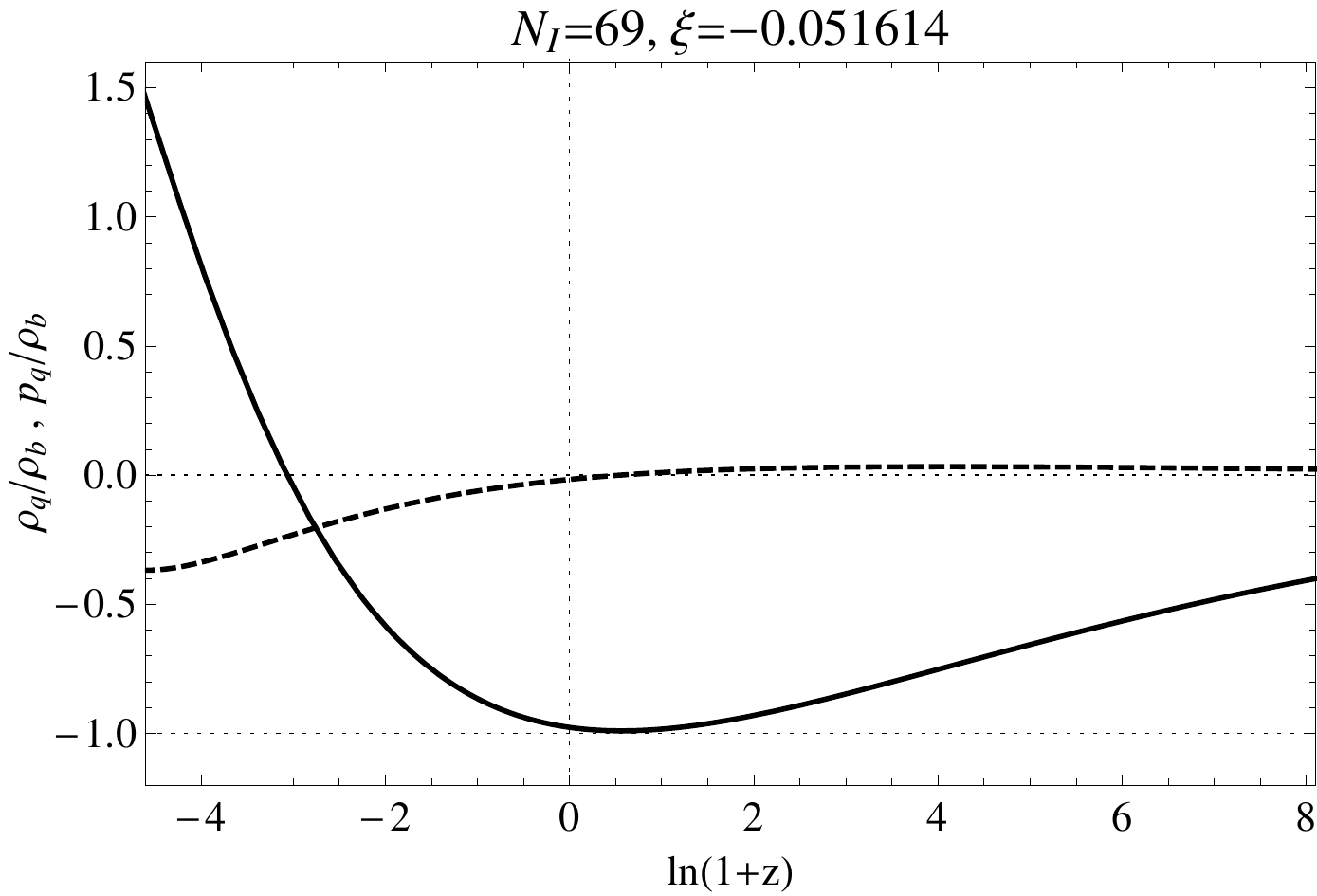}} \hfill \hfill
\subfloat[]{\includegraphics[width=8cm]{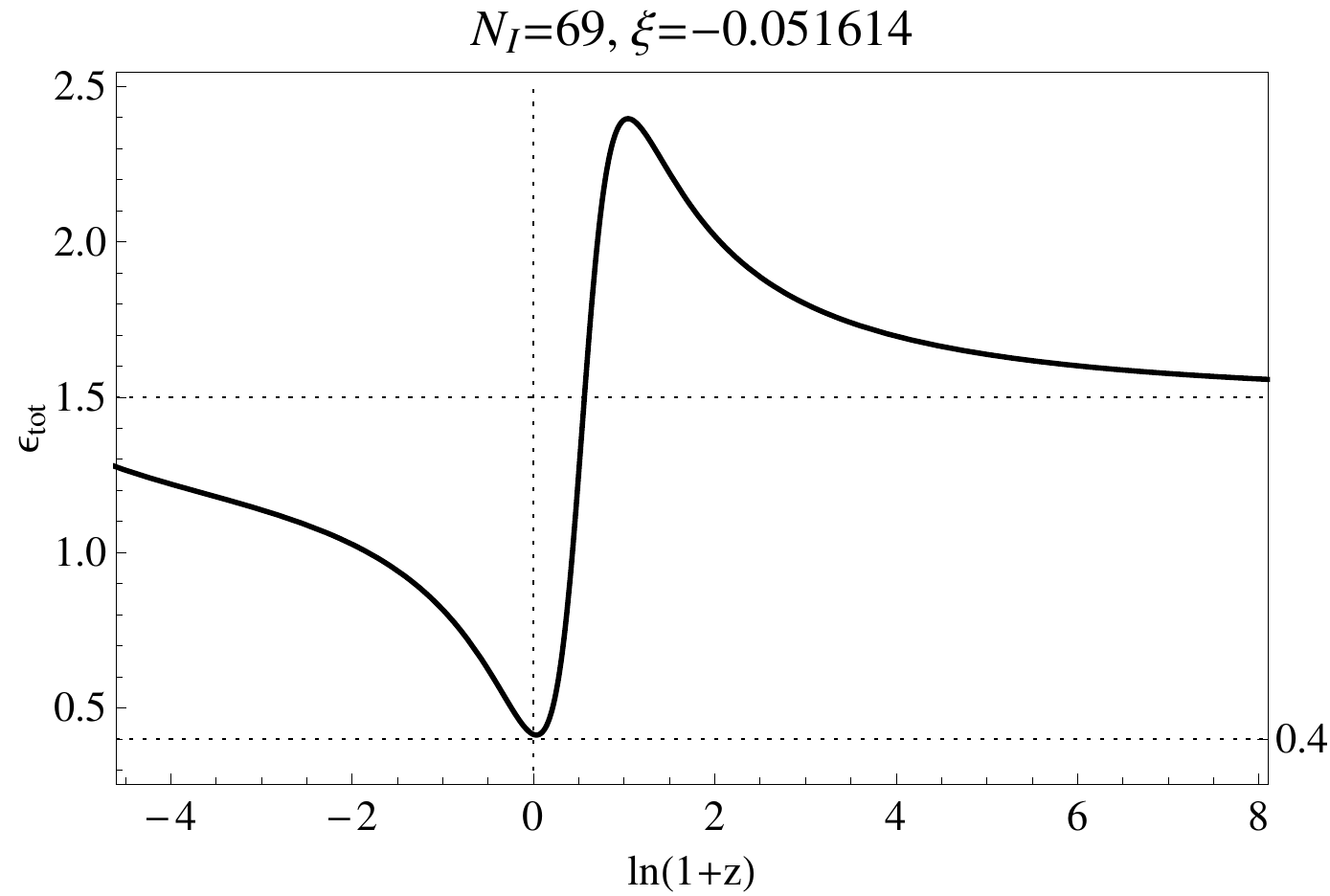}}

\caption{The two plots show 
(a) ratio of the quantum energy density (bold) and quantum pressure
(dashed) to the background energy density during matter dominated period, and
(b) naively corrected $\epsilon$ parameter during matter period,
both for choice of the number of e-foldings of inflation $N_I=69$
and non-minimal coupling $\xi=-0.051614$.}

\label{interesting}

\end{figure}

\vskip+2cm
\noindent There is (at least) one interesting choice of parameters,
$\xi=-0.051614$ and $N_I=69$, for which we are on the very edge of 
perturbative range up until today, for which the $\epsilon$ parameter 
takes the value $\epsilon \sim 0.4$, as it is measured today. The energy 
density, pressure, and $\epsilon$ parameter in this case are shown in
Figure \ref{interesting}.
For that choice of parameters at the 
end of inflation we have
$\rho_q/\rho_b\sim - 0.39$, and during radiation
$\rho_q/\rho_b\sim - 0.43$ which is mildly inside the perturbative regime,
but still large enough to affect the background evolution.

Of course, this estimate should not be taken too seriously. 
Backreaction in fact becomes too large to naively use (\ref{eps tot}).
It predicts that $\epsilon$ changes so much primarily because 
$\rho_q$ comes very close to $-\rho_b$
(but the direction of change is primarily determined
by the sign of the quantum fluid pressure). This means that the Hubble 
rate must decrease considerably (which can be inferred from 
the first Friedmann equation (\ref{friedmann1})), which is in
disagreement with the value measured today that we have assumed.
In order to find 
the true behavior, one would have to solve the one-loop quantum corrected
Friedmann equations self-consistently, much like in 
\cite{Suen:1987gu} (which is the only such study we are aware of). 
Nevertheless, this particular example indicates a very 
interesting possibility that this model, although finely tuned, could account
for the late-time acceleration of the Universe
(or other interesting possible effects), and certainly provides good
motivation for further investigation.

\section{Conclusions}
\label{sec:Conclusions}

In this paper we study the one-loop quantum backreaction of a 
non-minimally coupled massless scalar field (with vanishing
expectation value) on a universe that goes through a series of
constant $\epsilon$ eras connected by fast transitions. 
In our model the universe starts in a radiation era, which is followed by 
a period of inflation, radiation and finally a matter era 
(Figure \ref{expansion history}).
We assume a fixed, flat, FLRW background
metric and compute the one-loop expectation value of the 
energy-momentum tensor for our scalar in order to estimate its effect on
the background evolution. Our model is strictly speaking predictive
only when the quantum fluid is subdominant to the background. 
When the quantum backreaction becomes comparable to the background, 
a self-consistent
solution has to be found by performing a self-consistent treatment of the 
background equation in presence of the (one-loop) quantum fluctuations.
Such a treatment is left for future work, here we determine in which cases
the self-consistent treatment might yield interesting predictions.

\pagebreak

 Our main results and observations can be summarized as follows:
\begin{itemize}

\item[$\bullet$] When the  non-minimal coupling is negative, the quantum 
backreaction is negative and,
for a sufficiently large negative coupling, it grows with 
respect to the background, becoming eventually dominant (see e.g.~\cite{Janssen:2009nz}). 
This latter case can be treated by solving self-consistently the equations of semi-classical gravity.
For a minimal inflation with $N_{\rm efolds}\sim 66$, the value of $\xi$ 
for which the backreaction becomes significant at the end of inflation is about
$\xi\simeq -0.057$. It is plausible that such a backreaction would provide a graceful exit from inflation
without having to resort to fine tuning of the inflaton potential 
(such to give a vanishing contribution to the cosmological constant) at the end of inflation.
This question deserves further investigation. Furthermore, 
studying quantum backreaction during inflation can be used to constrain the physically acceptable values of 
non-minimal coupling for light scalar fields during inflation (adiabatic inflaton condensates 
such as the Higgs field of Higgs inflation are exempt from this constraint and they require a separate analysis). 

\item[$\bullet$] Similarly as found in the case of a 
minimally coupled scalar~\cite{Glavan:2013mra}, 
during radiation era the backreaction contribution of a non-minimally coupled scalar does not change significantly 
when compared with the background. This also means that, if the backreaction is subdominant
at the end of inflation, it will remain so at the beginning of matter era.

\item[$\bullet$] When $\xi<-1/3$, the relative quantum backreaction during matter era is negative and it grows.
However, for realistic duration of inflation ($N_I>65$), it already dominates during inflation,
making this case physically irrelevant. One can nevertheless construct a viable (albeit more complex) model in this case,
by making $\xi$ field dependent, $\xi=\xi(\chi)$ ($\chi$ symbolizing 
some other field $\phi$ is coupled to), and arranging it such that during inflation 
$\xi$ is above the critical value (for which quantum backreaction remains subdominant throughout inflation),
and $\xi<-1/3$ during matter era, such that the relative backreaction grows in matter era. It would be of interest to study 
the self-consistent evolution of the universe in this case. Namely, a negative backreaction slows down the rate of 
universe's expansion, and in the extreme case when it is large and negative, it could even reverse the expansion to a 
contraction, thus completely changing the future of our universe.

\item[$\bullet$] A potentially very interesting transient occurs for inflation 
close to minimal ($N_{\rm efolds}\sim 69$)
and the non-minimal coupling slightly larger than its critical value,
$\xi\sim-0.052\gtrsim\xi_c\simeq -0.057$. In this case,
the quantum backreaction exhibits a transient behaviour in recent 
times (illustrated in Figure \ref{interesting}),
and changes from negative to positive. This case bears resemblance to 
late time dark energy, and a self-consistent
treatment of semiclassical gravity is needed in order to find out
whether this is a good model for dark energy. 
Since there are essentially no free parameters in this model (the values of
the number of e-folds and the non-minimal coupling are to a large extent fixed 
by the requirement that the transient occurs in recent times and by the requirement that the backreaction reaches 
a maximum value that is comparable to the background energy density), this model can be easily tested by astronomical 
observations of the Hubble rate as a function of the redshift, and by studying late time evolution of small 
perturbations on top of the homogeneous background. These studies are however beyond of the scope of this work.  

\item[$\bullet$] In the limit when $\xi\rightarrow 0$ (minimal coupling) 
and $\mathcal{H}_0\rightarrow 0$ (very long inflation), our results 
agree with those of 
~\cite{Glavan:2013mra}. In particular, 
by taking a careful limit $\xi\rightarrow 0$, we find 
the term in the relative quantum backreaction 
that grows logarithmically in time during radiation (Appendix C), 
that is absent in the case when $\xi<0$.  

\item[$\bullet$] Our late time quantum backreaction is dominated by the infrared fluctuations, and does not depend
on details of the transitions, provided the transitions are fast (faster than the Hubble rate at the relevant transition).
 However, if any of the transitions is slow (a slow transition can be modeled  
by e.g. a tangent hyperbolic change of the $\epsilon$ parameter, 
or as was done in \cite{Koivisto:2010pj}), 
the leading order late time quantum backreaction
will in general acquire a dependence on the transition rate.
 In this case we expect a reduced amplitude of the late time backreaction, but an identical scaling in time.
Hence, including the case of slow transition(s) is inessential, and we will not study it further here.

\item[$\bullet$] It would be of interest to extend our analysis to inflationary fluctuations of other quantum fields.
Gravitons were already studied in~\cite{Janssen:2007ht} and~\cite{Glavan:2013mra}, and their contribution scales 
as matter in matter era, and constitute a small contribution to dark matter. A heuristic analysis suggests 
that a similar conclusion can be reached for adiabatic scalar cosmological perturbations~\cite{Glavan:2013mra},
although the case of a nonminimally coupled inflaton is more delicate and it requires a separate investigation.

\item[$\bullet$] Na\^ively, one might expect no significant quantum backreaction from
massless gauge fields such as photons. Indeed,   
at the classical level, photons couple conformally to gravity and hence their quantum backreaction
can be treated as that of any conformally coupled field. The quantum backreaction of conformal fields can be 
represented by local higher-derivative curvature terms, and a general study~\cite{Koksma:2008jn} 
shows that their late time backreaction is completely negligible.
However, when one includes couplings of gauge fields to other fields such as (light) massless scalars,
a significant photon production during inflation is in fact
possible~\cite{Prokopec:2002jn,Prokopec:2002uw}, making a detailed study of the late-time backreation
from gauge fields worthwhile.

\item[$\bullet$] One might think that the Pauli exclusion principle
will prevent fermions from giving a large quantum backreaction (because population of the infrared
states is limited to one fermion per state). This is however not true, since fermion loops contribute
with an opposite
sign, and hence can destabilise the vacuum of a scalar field in an expanding space-time
(for a study of a Yukawa theory on de Sitter case see Ref.~\cite{Miao:2006pn}).
This suggests that studying late time quantum backreaction from fermions
could yield interesting results for cosmology.

\item[$\bullet$] One might argue that the late time backreaction we calculate here is unobservable,
since it primarily comes from super-Hubble modes. This is however not true for the following reasons.
Firstly, the one-loop stress energy tensor we calculate is gauge invariant.~\footnote{
In order to see that, note that our scalar field is a spectator field,
and thus its expectation value is zero in
all epochs, i.e. we have, $\hat \Phi(x) = \phi_0(t) + \hat\varphi(x)$, where
$\langle\Omega|\hat\Phi(x)|\Omega\rangle = \phi_0(t) \rightarrow 0$. This then implies
that the stress energy tensor is gauge invariant (to quadratic order in perturbations). This
can be shown as follows. Under coordinate transformations $x^\mu\rightarrow x^\mu+\xi^\mu(x)$,
a scalar field transform as, $\Phi(x)\rightarrow \tilde\Phi(x)=\Phi(x)-x^\mu\partial_\mu\tilde\Phi(x)$
plus higher order terms. In the absence of a condensate $\phi_0$, this reduces to,
$\varphi(x)\rightarrow \tilde\varphi(x)=\varphi(x)-x^\mu\partial_\mu\varphi(x)$ plus cubic order
terms (here we took $\xi^\mu$ and $\varphi$ to be first order quantities). The one-loop stress energy
tensor can be represented as a differential operator acting on
$\langle\Omega|\hat\varphi(x)^2|\Omega\rangle$,
which to lowest order transforms as, $\langle\Omega|\hat\varphi(x)^2|\Omega\rangle\rightarrow
\langle\Omega|\hat{\tilde\varphi}(x)^2|\Omega\rangle=\langle\Omega|\hat\varphi(x)^2|\Omega\rangle
-\xi^\mu\partial_\mu \langle\Omega|\hat\varphi(x)^2|\Omega\rangle$ plus higher order terms.
Notice that the leading correction is cubic in perturbations, and hence
$\langle\Omega|\hat\varphi(x)^2|\Omega\rangle$ is gauge invariant to quadratic order in
gauge transformations. which completes the proof.}
And secondly, its value is invariant under spatial translations
(this imme-\vskip+2cm
\noindent diately  follows from the fact that $T_{\mu\nu}^q$ depends on time but it is
 independent of  space), and hence each local (freely falling)
observer on a FLRW background will agree on the value of the backreaction. 
Our calculation neglects small fluctuations on top of a homogeneous expanding background.
In order to study how these fluctuations affect the result presented here, one would have to study
$\langle\Omega|\hat T_{\mu\nu}(x)\hat T_{\rho\sigma}(x')|\Omega\rangle$, which is worth doing
once one has a good candidate for dark energy.

\end{itemize}

\section*{Acknowledgments}

This work is part of the D-ITP consortium, a program of the Netherlands
Organisation for Scientific Research (NWO) that is funded by the Dutch
Ministry of Education, Culture and Science (OCW).

\appendix

\section{Renormalization of the energy-momentum tensor on smooth FLRW}
\label{sec:Renormalization}

This appendix presents an outline of the regularization and renormalization 
procedures needed to assign finite values to expressions (\ref{rho def}) 
and (\ref{p def}). The exposition is essentially the same as in Appendix A from 
\cite{Glavan:2013mra}, just extended to include non-minimal coupling.
The material here is standard (see \cite{Birrell:1982ix}), and we include it
for the sake of notation and completeness. For a different method of
regularization see \cite{Weinberg:2010wq}. We think that removing the 
depndence on the UV cutoff is essential for obtaining reliable results,
since it can lead to some spurious time evolution as in the case of
e.g. just introducing a comoving momentum cutoff \cite{Albareti:2014ria}, or
introducing a physical momentum cutoff \cite{Sheikhahmadi:2014rka} which 
also leads to non-conservation of the energy-momentum tensor.

\vskip+0.4cm

Integrals in (\ref{rho def}) and (\ref{p def}) are split into UV and IR parts
by some constant cut-off scale $\mu\gg\mathcal{H}$. The
IR parts can be evaluated in $D=4$ right away. The UV part needs to be
regularized, and the method of choice is dimensional regularization. It 
entails evaluating the integrals in arbitrary $D$ dimensions for which
the integrals converge. This automatically eliminates any power-law divergences
in $D=4$, but not the logarithmic one, which in dimensional regularization
presents itself as $1/(D-4)$ term. That term is to be absorbed into the
higher-derivative gravitational counterterms and after that the limit $D\rightarrow4$
is to be taken. This entails renormalization after which we are left with a finite
final answer for energy density and pressure.

The practical task is to isolate this $1/(D-4)$ divergence and show that it can be
absorbed into the counterterm. The integrals to evaluate are
\begin{align}
\rho_q^{UV,0} ={}& \frac{a^{-D}}{(4\pi)^{\frac{D-1}{2}}
	\Gamma\left( \frac{D-1}{2} \right)} \int_{\mu}^{\infty} dk \, k^{D-2}
	\Bigg{[} 2 k^2 |u|^2 - \frac{1}{2}[D-2-4\xi(D-1)] \mathcal{H}' |u|^2
\nonumber \\
&	\qquad - \frac{1}{2}[D-2-4\xi(D-1)]\mathcal{H}
	\frac{\partial}{\partial\eta}|u|^2
	+ \frac{1}{2} \frac{\partial^2}{\partial\eta^2}|u|^2 \Bigg{]} , 
	\label{rhoQ UV}\\
p_q^{UV,0} ={}& \frac{a^{-D}}{(4\pi)^{\frac{D-1}{2}}
	\Gamma\left( \frac{D-1}{2} \right)} \int_{\mu}^{\infty} dk \, k^{D-2}
	\Bigg{[} \frac{2k^2}{D-1} |u|^2 - \frac{1}{2}[D-2-4\xi(D-1)] 
	\mathcal{H}' |u|^2
\nonumber \\
&	\qquad - \frac{1}{2}[D-2-4\xi(D-1)]\mathcal{H}
	\frac{\partial}{\partial\eta}|u|^2
	+ \frac{1}{2}(1-4\xi) \frac{\partial^2}{\partial\eta^2}|u|^2 \Bigg{]} .
	\label{pQ UV}
\end{align}
This is to be done by finding the UV asymptotic expansion of the mode function,
integrating term by term, keeping the terms that are divergent or finite in
the limits $D\rightarrow4$ and $\mu\rightarrow\infty$, and throwing away the ones
that are zero in this limit.

The asymptotic expansion is obtained as a Wentzel-Kramers-Brillouin (WKB)
approximation of the equation of motion (\ref{EOM}), since 
$k\ge\mu\gg\mathcal{H}$.
The shortcut way of doing this is to assume the positive-frequency
expansion of $u$ in powers of $1/k$,
\begin{equation} \label{u expansion}
u = \frac{e^{-ik\eta}}{\sqrt{2k}} \left[ 1 + \frac{iF_1(\eta)}{k}
	+ \frac{F_2(\eta)}{k^2} + \frac{iF_3(\eta)}{k}
	+ \frac{F_4(\eta)}{k^4} \right] + \mathcal{O}(k^{-5}) .
\end{equation}
We have the freedom to multiply this mode function by a constant phase factor,
but since it is physically irrelevant, here we chose the most convenient one for
the discussion at hand.
Now we can solve for $F_n(\eta)$'s by solving the equation of motion (\ref{EOM})
order by order in $1/k$,
\begin{align}
F_1' ={}& - \frac{1}{2}f , \label{F1} \\
F_2' ={}& \frac{1}{2} \Big{[} F_1'' + f F_1 \Big{]} , \label{F2} \\
F_3' ={}& -\frac{1}{2} \Big{[} F_2'' + f F_2 \Big{]} , \label{F3} \\
F_4' ={}& \frac{1}{2} \Big{[} F_3'' + f F_3 \Big{]} , \label{F4}
\end{align}
where $f$ is defined in (\ref{f}), and the same for Wronskian normalization 
condition,
\begin{align}
& 2 F_2 + F_1^2 - F_1' = 0 , \label{W1} \\
& 2 F_4 + 2 F_3F_1 + F_2^2 - F_3' + F_2'F_1 - F_1F_2 = 0 . \label{W2}
\end{align}
Integrating equations (\ref{F1})-(\ref{F4}) and imposing conditions (\ref{W1}), 
(\ref{W2}) yields
\begin{align}
F_1(\eta) ={}& - \frac{1}{2} \int_{\eta_0}^{\eta} d\tau\, f(\tau) , \\
F_2(\eta) ={}& - \frac{f(\eta)}{4} 
	- \frac{1}{8} \left[ \int_{\eta_0}^{\eta} d\tau\, f(\tau) \right]^2 , \\
F_3(\eta) ={}& \frac{1}{8}[f'(\eta)-f'(\eta_0)]
	+ \frac{f(\eta)}{8} \int_{\eta_0}^{\eta} d\tau \, f(\tau)
	+ \frac{1}{48}\left[ \int_{\eta_0}^{\eta} d\tau\, f(\tau) \right]^3
	+ \frac{1}{8} \int_{\eta_0}^{\eta} d\tau\, f^2(\tau) , \\
F_4(\eta) ={}& \frac{f''(\eta)}{16} + \frac{5 f^2(\eta)}{32} 
	+ \frac{1}{16}[f'(\eta)-f'(\eta_0)] \int_{\eta_0}^{\eta}d\tau f(\tau)
	+ \frac{f(\eta)}{32} \left[ \int_{\eta_0}^{\eta} d\tau f(\tau) \right]^2
\nonumber \\
&	\quad + \frac{1}{384} \left[ \int_{\eta_0}^{\eta} d\tau \, f(\tau) \right]^4
	+ \frac{1}{16} \left[ \int_{\eta_0}^{\eta} d\tau\, f(\tau) \right]
		\left[ \int_{\eta_0}^{\eta} d\tau\, f^2(\tau) \right] ,
\end{align}
where we have imposed the initial condition that all $F_n$'s for odd $n$'s are
zero at $\eta_0$. This, again, is nothing other than picking an arbitrary phase.

An important thing to note here is that the time evolution 
of the mode function in the
UV is adiabatic, it does not change its positive frequency character. 
In other words, the evolution will induce no mode mixing in the UV to any adiabatic 
order, at most it will be suppressed to decay faster in the UV than any power
of $1/k$. This is quantified as a result on the Bogolyubov coefficients,
$\alpha(k)\overset{k\rightarrow\infty}{\rightarrow}1$ and 
$\beta(k)\overset{k\rightarrow\infty}{\rightarrow}0$ faster than any power of $1/k$.

Now it is a simple matter to evaluate (\ref{rhoQ UV}) and (\ref{pQ UV}),
expand the result around $D=4$ and discard the terms $\mathcal{O}(D-4)$
and $\mathcal{O}(\mu^{-1})$,
\begin{align}
\rho_q^{UV,0} ={}& \frac{1}{16\pi^2a^4} \Bigg{\{} - \mu^4
	- (1-6\xi)\mathcal{H}^2\mu^2
	- \frac{1}{2}(1-6\xi)^2 \Big{[} -2\mathcal{H}''\mathcal{H}
		+ (\mathcal{H}')^2 + 3\mathcal{H}^4 \Big{]} 
		\frac{\mu^{D-4}}{D-4} 
\nonumber \\
&	 + \frac{1}{2} \Big{[} (1-6\xi)^2 \ln a - c  \Big{]} 
	\Big{[} -2\mathcal{H}''\mathcal{H}
		+ (\mathcal{H}')^2 + 3\mathcal{H}^4 \Big{]}
	- \frac{3}{2}(1-6\xi)^2\mathcal{H}^4 \Bigg{\}} ,
\label{rho exp UV}
\\
p_q^{UV,0} ={}& \frac{1}{48\pi^2a^4} \Bigg{\{} - \mu^4
	+ (1-6\xi)(2\mathcal{H}' - \mathcal{H}^2)\mu^2 
\nonumber \\
&	- \frac{1}{2} \Big{[} 2\mathcal{H}''' - 2\mathcal{H}''\mathcal{H}
		+ (\mathcal{H}')^2 - 12 \mathcal{H}'\mathcal{H}^2
		+ 3\mathcal{H}^4 \Big{]} \frac{\mu^{D-4}}{D-4}
\nonumber \\
&	+ \frac{1}{2} \Big{[} (1-6\xi)^2 \ln a - c  \Big{]} \Big{[} 
		2\mathcal{H}''' - 2\mathcal{H}''\mathcal{H}
		+ (\mathcal{H}')^2 - 12\mathcal{H}'\mathcal{H}^2
		+ 3 \mathcal{H}^4 \Big{]}
\nonumber \\
&	+ \frac{1}{6}(1-6\xi)^2 \Big{[} 2\mathcal{H}''' 
		- 2\mathcal{H}''\mathcal{H} + (\mathcal{H}')^2
		+ 24 \mathcal{H}'\mathcal{H}^2 - 6\mathcal{H}^4 \Big{]} 
	\Bigg{\}} ,
	\label{p exp UV}
\end{align}
where $c = (1-6\xi) [\frac{1}{2}(1-6\xi)(\gamma_E-\ln\pi)+2\xi]$, 
and $\gamma_E$ is the Euler-Mascheroni constant.

The action for the counterterm needed to absorb the logarithmic divergence is
\begin{equation}
S_{ct} = \alpha_{ct} S_1 = \alpha_{ct} \int d^Dx \sqrt{-g}\, R ,
\end{equation}
where
\begin{equation}
\alpha_{ct} = \frac{1}{1152\pi^2}
	\left[ (1-6\xi)^2 \left( \frac{\mu^{D-4}}{D-4} \right)
	+ \alpha_f \right] ,
\end{equation}
and $\alpha_f$ is a free finite constant (that can depend on $\mu$ logarithmically),
to be determined by measurements in principle. $S_{ct}$ gives the following
contribution to the energy-momentum tensor
\begin{equation}
\alpha_{ct}\times {}^{(1)}H_{\mu\nu}
	= \alpha_{ct} \times \frac{(-2)}{\sqrt{-g}}\frac{\delta S_1}{\delta g^{\mu\nu}}
	= \alpha_{ct} \Big{(} 4\nabla_\mu \nabla_\nu R
		- 4g_{\mu\nu} \square R + g_{\mu\nu} R^2 
		- 4 R_{\mu\nu} R \Big{)} .
\end{equation}
For FLRW space-time ${}^{(1)}H_{\mu\nu}$ is diagonal, and its expansion
around $D=4$ reads
\begin{align}
\rho_q^{ct} = \frac{\alpha_{ct}}{a^2} \times {}^{(1)}H_{00}
	={}& \frac{(1-6\xi)^2}{32\pi^2a^4} \Big{[} -2\mathcal{H}''\mathcal{H}
		+ (\mathcal{H}')^2 + 3\mathcal{H}^4 \Big{]} 
		\left( \frac{\mu^{D-4}}{D-4} \right)
\nonumber \\
&	 + \frac{(1-6\xi)^2}{96\pi^2a^4} \Big{[} -4\mathcal{H}''\mathcal{H}
		+ 2(\mathcal{H}')^2 - 6\mathcal{H}'\mathcal{H}^2
		+ 9\mathcal{H}^4 \Big{]}
\nonumber \\
&	+ \frac{\alpha_f}{32\pi^2a^4} \Big{[} -2\mathcal{H}''\mathcal{H}
		+ (\mathcal{H}')^2 + 3\mathcal{H}^4 \Big{]} ,
\\
\delta_{ij }p_q^{ct} = \frac{\alpha_{ct}}{a^2} \times {}^{(1)}H_{ij}
	={}& \delta_{ij} \frac{(1-6\xi)^2}{96\pi^2a^4}
		\Big{[} 2\mathcal{H}''' - 2\mathcal{H}''\mathcal{H}
		+ (\mathcal{H}')^2 - 12\mathcal{H}'\mathcal{H}^2
		- 3\mathcal{H}^4 \Big{]} \left( \frac{\mu^{D-4}}{D-4} \right)
\nonumber \\
& 	+ \delta_{ij} \frac{(1-6\xi)^2}{288\pi^2a^4} \Big{[} 
		2\mathcal{H}''' + 10\mathcal{H}''\mathcal{H}
		+10(\mathcal{H}')^2 - 30\mathcal{H}'\mathcal{H}^2
		- 3 \mathcal{H}^4 \Big{]}
\nonumber \\
&	+ \delta_{ij} \frac{\alpha_f}{96\pi^2a^4} \Big{[} 
		 2\mathcal{H}''' - 2\mathcal{H}''\mathcal{H}
		+ (\mathcal{H}')^2 - 12\mathcal{H}'\mathcal{H}^2
		- 3\mathcal{H}^4 \Big{]} .
\end{align}
The terms divergent in $D=4$ in the two expressions above cancel the 
divergent terms in (\ref{rho exp UV}) and (\ref{p exp UV}).

There is one more contribution to the energy-momentum tensor, which
survives even in the conformal limit $\xi\rightarrow 1/6$, the so-called
conformal anomaly \cite{Capper:1974ic, Duff:1977ay, Birrell:1982ix}.
 It is not strictly necessary
for renormalization on FLRW space-time, but it is on more general ones, so 
we include it here. Its contribution on FLRW is
\begin{align}
\rho_q^{ca} ={}& \frac{1}{2880\pi^2a^4} \Big{[} 2\mathcal{H}''\mathcal{H}
		- (\mathcal{H}')^2 \Big{]} + \frac{\alpha_{ca}}{32\pi^2a^4}
	\Big{[} 2\mathcal{H}''\mathcal{H} - (\mathcal{H}')^2
		- 3\mathcal{H}^4 \Big{]} , \\
p_q^{ca} ={}& - \frac{1}{8640\pi^2a^4} \Big{[} 2\mathcal{H}'''
		- 2\mathcal{H}''\mathcal{H} + (\mathcal{H}')^2 \Big{]}
\nonumber \\
&	\qquad	+ \frac{\alpha_{ca}}{96\pi^2a^4} \Big{[} -2\mathcal{H}''' 
		+ 2\mathcal{H}''\mathcal{H} - (\mathcal{H}')^2
		+ 12 \mathcal{H}'\mathcal{H}^2 - 3\mathcal{H}^4 \Big{]} ,
\end{align}
where $\alpha_{ca}$ is a free constant.

The final answer for the UV part of the energy-momentum tensor is obtained
by adding all these three contributions together,
\begin{align}
\rho_q^{UV} ={}& \rho_q^{UV,0} + \rho_q^{ct} + \rho_q^{ca}
\nonumber \\
={}& \frac{1}{16\pi^2a^4} \Big{[} - \mu^4 - (1-6\xi)\mathcal{H}\mu^2 \Big{]}
	+ \frac{(1-6\xi)^2}{32\pi^2a^4} \Big{[} -2\mathcal{H}''\mathcal{H}
	+ (\mathcal{H}')^2 + 3\mathcal{H}^4 \Big{]} (\ln a + \widetilde{\alpha}_f)
\nonumber \\
&	+ \frac{(1-6\xi)^2}{48\pi^2a^4} \Big{[} -2\mathcal{H}''\mathcal{H}
		+ (\mathcal{H}')^2 - 3\mathcal{H}'\mathcal{H}^2 \Big{]}
	+ \frac{1}{2880\pi^2a^4} \Big{[} 2\mathcal{H}''\mathcal{H}
		- (\mathcal{H}')^2 \Big{]} ,
\\
p_q^{UV} ={}& p_q^{UV,0} + p_q^{ct} + p_q^{ca}
\nonumber \\
={}& \frac{1}{48\pi^2a^4} \Big{[} \mu^4 
		+(1-6\xi)(2\mathcal{H}'-\mathcal{H}^2)\mu^2 \Big{]}
\nonumber \\
&	+ \frac{(1-6\xi)^2}{96\pi^2a^4} \Big{[} 2\mathcal{H}'''
	- 2\mathcal{H}''\mathcal{H} + (\mathcal{H}')^2
	- 12 \mathcal{H}'\mathcal{H}^2 + 3\mathcal{H}^4 \Big{]} 
		(\ln a+\widetilde{\alpha}_f)
\nonumber \\
& 	+ \frac{(1-6\xi)^2}{288\pi^2a^4} \Big{[} 4\mathcal{H}'''
		+ 8 \mathcal{H}''\mathcal{H} + 11 (\mathcal{H}')^2
		- 6\mathcal{H}'\mathcal{H}^2 - 9\mathcal{H}^4 \Big{]}
\nonumber \\
&	- \frac{1}{8640\pi^2a^4} \Big{[} 2\mathcal{H}'''
		- 2\mathcal{H}''\mathcal{H} + (\mathcal{H}')^2 \Big{]} ,
\end{align}
where all the free constants $c$, $\alpha_f$ and $\alpha_{ca}$ 
have been collected in $\widetilde{\alpha}_f$
since they multiply the same contributions. Energy density and pressure 
given above satisfy the conservation equation (\ref{conservation eq})
by themselves. Specializing to FLRW backgrounds with constant $\epsilon$
parameter, the UV contributions to the energy density and pressure are
\begin{align}
\rho_q^{UV} ={}& \frac{1}{16\pi^2a^4} \Bigg{\{}
	-\mu^4 - (1-6\xi) \mathcal{H}^2\mu^2 
	+ \frac{3}{2}(1-6\xi)^2 \epsilon(2-\epsilon) \mathcal{H}^4
		(\ln a + \widetilde{\alpha}) 
\nonumber \\
&	- (1-6\xi)^2 (1-\epsilon)(2-\epsilon) \mathcal{H}^4 
	+ \frac{1}{60}(1-\epsilon)^2 \mathcal{H}^4 \Bigg{\}} ,
\\
p_q^{UV} ={}& \frac{1}{48\pi^2a^4} \Bigg{\{} 
	- \mu^4 + (1-6\xi)(1-2\epsilon) \mathcal{H}^2\mu^2
	- \frac{3}{2}(1-6\xi)^2 \epsilon(2-\epsilon)(3-4\epsilon)\mathcal{H}^4
		(\ln a + \widetilde{\alpha})
\nonumber \\
&	+ \frac{1}{2}(1-6\xi)^2(2-\epsilon)(6-17\epsilon+8\epsilon^2)
		\mathcal{H}^4
	- \frac{1}{60}(1-\epsilon)^2(3-4\epsilon)\mathcal{H}^4 \Bigg{\}} .
\end{align}
The full answer for the one-loop expectation value for energy density and pressure
is obtained by adding up the UV and IR contributions and taking the limit
$\mu\rightarrow\infty$,
\begin{equation}
\rho_q = \lim_{\mu\rightarrow\infty} 
	\Big{(} \rho_q^{UV} + \rho_q^{IR} \Big{)},
	\qquad p_q = \lim_{\mu\rightarrow\infty} 
	\Big{(} p_q^{UV} + p_q^{IR} \Big{)} .
\end{equation}

\section{Cosmological parameters}
\label{parameters}

This appendix serves as a reminder of how to relate quantities
$\mathcal{H}_i$ we have used in the calculation with the values of cosmological
parameters usually used. We basically use relation (\ref{HinA}),
\begin{equation}
\mathcal{H}_{i}=\mathcal{H}_{i-1} \left( \frac{a_i}{a_{i-1}} 
		\right)^{1-\epsilon_i}
	= \mathcal{H}_{i-1} \, e^{(1-\epsilon_i) N_i } ,
\end{equation}
to accomplish this, where. $N_i=\ln(a_i/a_{i-1})$ is the number of e-foldings
of the $i$th period.
The cosmological parameters (except $\epsilon_I$) are taken from 
\cite{Ade:2014xna}. By taking the Hubble rate today (with $a_{tod.}=1$) to be
\begin{equation}
H_{tod.}=68\, \mathrm{\frac{km}{Mpc\, s}} = \mathcal{H}_{tod} ,
\end{equation}
and the redshift of radiation-matter equality,
\begin{equation}
z_{eq} = 3270 = \frac{1}{a_2}+1 ,
\end{equation}
we get that the (conformal) Hubble rate at the time of transition
between radiation and matter period is
\begin{equation}
\mathcal{H}_2 = 3888\, \mathrm{\frac{km}{Mpc\, s}} ,
\end{equation}
and the duration of matter period expressed in the number of e-foldings
\begin{equation}
N_M = 8.09 .
\end{equation}

The condition of minimal inflation that the conformal Hubble rate 
at the beginning of inflation be equal to the one today,
$\mathcal{H}_{0,min}=\mathcal{H}$, gives one condition on the e-foldings,
\begin{equation} \label{cond1}
0 = (1-\epsilon_I)N_{I,min} - N_R - \frac{1}{2}N_M .
\end{equation}
The other condition needed is supplied by the amplitude of the scalar 
perturbations in the CMB. It fixes the physical Hubble rate
\begin{equation}
H_* = 3.45\times 10^{13}\, \mathrm{\frac{GeV}{\hbar}}
	= 1.618\times 10^{60}\, {\frac{\text{m}}{\text{Mpc s}}}
\end{equation}
 at the time
the mode at the pivotal scale $k_*=0.002\, \mathrm{Mpc}^{-1}$ 
left the horizon during inflation. For minimal inflation the Hubble scale
is the same as today, $k_0=0.00026\, \mathrm{Mpc}^{-1}$,
which gives us
\begin{equation}
\frac{k_*}{k_0} = \frac{\mathcal{H}_*}{\mathcal{H}_{0,min}} 
	= \left( \frac{a_*}{a_{0,min}} \right)^{1-\epsilon_I} .
\end{equation}
Using the minimal inflation assumption $H_{tod.}=\mathcal{H}_0$, this condition
can be rewritten as
\begin{equation} \label{cond2}
N_{I,min} + N_R + N_R = \ln\left[ \frac{H_*}{H_{tod.}}
	\left( \frac{k_0}{k_*} \right)^{\frac{\epsilon_I}{1-\epsilon_I}} \right] .
\end{equation}
Results from \cite{Ade:2014xna} would suggest that $\epsilon_I=0.01$, which,
together with the two conditions (\ref{cond1}) and (\ref{cond2}) fixes the 
duration of radiation period and minimal inflation,
\begin{equation}
N_R = 60.77 , \qquad N_{I,min} = 65.47 , 
\end{equation}
from where we can infer the Hubble rate at the end of inflation
(at the beginning of inflation it is the same as today)
\begin{equation}
\mathcal{H}_1 = 9.59\times 10^{29}\, \frac{\text{km}}{\text{Mpc s}} .
\end{equation}
%




\section{Backreaction during radiation period}

In this appendix we repeat the calculation from Section VIII, but this time for
the backreaction during radiation period (Figure \ref{expansion history}).
The hierarchy of scales assumed is
$\mathcal{H}_0\ll \mathcal{H}\ll\mathcal{H}_1$. This is strictly valid only some 
time after the transition from inflation to radiation period when the conformal
Hubble rate has dropped enough. 

There is no need to repeat here step by step all the approximation steps
for Bogolyubov coefficients from Section IX.
The approximation for $Z_{Bog.}$ here is obtained by making substitutions
$A_{3,1} \rightarrow A_{2,1}$, $\nu_I\rightarrow \nu_R=\frac{1}{2}$, and
$u_M\rightarrow u_R$ in the expression (\ref{Z approx}),
\begin{equation}\label{Z rad}
Z_{Bog.}(k,\eta) = \frac{8 |A_{2,1}|^2}{k^{2+2\nu_I}}
	\left[ \frac{1}{2}\frac{\partial^2}{\partial\eta_0^2} 
		+ 2k^2 + f_I(\eta_0) \right]
	 \Re^2[u_I(k,\eta_0)] \Re^2[u_R(k,\eta)] ,
\end{equation}
where
\begin{equation}
|A_{2,1}|^2 = \frac{\Gamma^2(\nu_I)(\nu_I-\frac{1}{2})^2}{16\pi}
	[2(1-\epsilon_I)\mathcal{H}_1]^{2\nu_I+1} .
\end{equation}
The integrals (\ref{J integral}) for $J_s$ to be evaluated here are
\begin{align} \label{Js radiation}
\mathcal{J}_s ={}& 
	\int\limits_{0}^{\mu}\! dk\, k^{2s-1-2\nu_I-2\nu_M} \,
	\Re^2[u_I(k,\eta_0)] \, \Re^2[u_R(k,\eta)]
\nonumber \\
={}& \frac{\pi}{8(1-\epsilon_I)\mathcal{H}_0}
	\int\limits_{0}^{\mu}\! dk\, k^{2s-3-2\nu_I} \,
	J_{\nu_I}^2\left( \frac{k}{(1-\epsilon_I)\mathcal{H}_0} \right)
	\sin^2 \left( \frac{k}{\mathcal{H}} \right)  , \qquad s=1,2,3 ,
\end{align}
and the solution for them is ($\mu$-independent term in the $1/\mu$ expansion)
\begin{align}
\mathcal{J}_s ={}& \frac{\sqrt{\pi}\ \Gamma(s-1)\Gamma(\frac{3}{2}-s+\nu_I)}
	{32\, \Gamma(2-s+\nu_I)\Gamma(2-s+2\nu_I)} 
	[(1-\epsilon_I)\mathcal{H}_0]^{-3+2s-2\nu_I}
\nonumber \\
&	\qquad \times \left[ 1 - {}_3F_2\left( -1+s, -1+s-2\nu_I,-1+s-\nu_I;
		\frac{1}{2}, -\frac{1}{2}+s-\nu_I ; 
		\frac{(1-\epsilon_I)^2\mathcal{H}_0^2}{\mathcal{H}^2} \right) \right]
\nonumber \\
&	+ 2^{-1-2s+2\nu_I} \, \Gamma(-3+2s-2\nu_I) \sin[\pi(s-\nu_I)]
		\mathcal{H}^{-3+2s-2\nu_I}
\nonumber \\
&	\qquad \times {}_3F_2 \left( \frac{1}{2}, \frac{1}{2}-\nu_I,
		\frac{1}{2}+\nu_I; 2-s+\nu_I, \frac{5}{2}-s+\nu_I;
		\frac{(1-\epsilon_I)^2\mathcal{H}_0^2}{\mathcal{H}^2} \right) \ .
\end{align}
This can be further simplified by recognizing that during radiation period
$\mathcal{H}\gg\mathcal{H}_0$ is always satisfied, so we can expand in
$\mathcal{H}_0/\mathcal{H}$,
\begin{align}
\mathcal{J}_s \approx{}&  2^{-1-2s+2\nu_I} \, \Gamma(-3+2s-2\nu_I)
	\sin[\pi(s-\nu_I)] \mathcal{H}^{-3+2s-2\nu_I}
\nonumber \\
&	+ \frac{\sqrt{\pi}\,\Gamma(s)\Gamma(\frac{1}{2}-s+\nu_I)}
	{16\Gamma(1-s+\nu_I)\Gamma(1-s+2\nu_I)}
	\frac{[(1-\epsilon_I)\mathcal{H}_0]^{-1+2s-2\nu_I}}{\mathcal{H}^2} \ ,
\end{align}
where, depending on $s$ and $\nu_I$ the first or the second term might be
dominant. Now we can form $\mathcal{I}_s$ integrals easily via (\ref{IfromJ}),
and from (\ref{rhoQinI}) and (\ref{pQinI}) calculate the energy density
and pressure, which to leading order in $\mathcal{H}_0/\mathcal{H}$ are
\begin{align}
\rho_q ={}& \frac{3\mathcal{H}_0^4}{16\pi^2a^4} 
	\left( \frac{\mathcal{H}_1}{\mathcal{H}_0} \right)^{2\nu_I+1}
	\frac{\xi (1-6\xi) (2-\epsilon_I) (1-\epsilon_I)^2 (\nu_I-\frac{1}{2})^2}
		{(\nu_I-\frac{3}{2})} , \\
p_q ={}& \frac{\mathcal{H}_0^4}{16\pi^2a^4} 
	\left( \frac{\mathcal{H}_1}{\mathcal{H}_0} \right)^{2\nu_I+1}
	\frac{\xi (1-6\xi) (2-\epsilon_I) (1-\epsilon_I)^2 (\nu_I-\frac{1}{2})^2}
		{(\nu_I-\frac{3}{2})} .
\end{align}
The backreaction during radiation period behaves as an ideal fluid with a 
constant equation of state parameter
\begin{equation}
w_q = \frac{1}{3} = w_b ,
\end{equation}
which means it scales just like radiation that dominates the background.
Therefore, the ratio of the quantum to classical energy density during radiation
period is constant, and for the choice of parameters of interest from Section IX,
$N_I=69$ and $\xi=-0.051614$, this ratio is
\begin{equation}
\frac{\rho_q}{\rho_b} = - 0.4335 .
\end{equation}
Therefore, the dominant contribution to the energy density is negative, scales 
the same as the background, but is smaller than the background.

This result does not reproduce the $\epsilon_I\rightarrow0$, 
$\xi\rightarrow0$, $\mathcal{H}_0\rightarrow0$ limit of
\cite{Glavan:2013mra}. The reason is that in this limit some 
$\mu$-dependent terms become $\mu$-independent, and of the same
order as the leading term.
Fortunately, there is another way we can calculate this result, 
by evaluating simpler integrals, and we can reproduce the limit of
\cite{Glavan:2013mra}.

We go back to integrals $\mathcal{I}_s$ in (\ref{Is IR}) and split the integration
intervals by introducing another scale $\mu_0$ such that
$\mathcal{H}_0\ll\mu_0\ll\mathcal{H}\ll\mu\ll\mathcal{H}_1$,
\begin{equation}
\mathcal{I}_s^{(0,\mu_0)} + \mathcal{I}_s^{(\mu_0,\mu)} = 
	\left( \int_{0}^{\mu_0} + \int_{\mu_0}^{\mu} \right) dk\, k^{2s}
	Z_{Bog.}(k,\eta) .
\end{equation}
On the interval $(0,\mu_0)$ we can follow the same procedure
(\ref{Z rad})-(\ref{Js radiation}), just that the integration is up to $\mu_0$,
and we can further expand integral (\ref{Js radiation}) in $k/\mathcal{H}$ to get
\begin{align}
\mathcal{J}_s^{(0,\mu_0)} = {}&
	\frac{\pi}{8(1-\epsilon_I)\mathcal{H}_0\mathcal{H}^2}
	\int\limits_{0}^{\mu_0}\! dk\,k^{2s-1-2\nu_I} 
	J_{\nu_I}^2\left( \frac{k}{(1-\epsilon_I)\mathcal{H}_0} \right)
\nonumber \\
={}& \frac{2^{-4-2\nu_I} \pi \mu_0^{2s}
	 [(1-\epsilon_I)\mathcal{H}_0]^{-1-2\nu_I}}
	{s\, \Gamma^2(1+\nu_I) \mathcal{H}^2}
	{}_2F_3\left( s, \frac{1}{2}+\nu_I; 1+s, 1+\nu_I, 1+2\nu_I
	; \frac{-\mu_0^2}{(1-\epsilon_I)^2\mathcal{H}_0^2} \right)
\nonumber \\
={}& \frac{\sqrt{\pi}\, \Gamma(s)\Gamma(\frac{1}{2}-2+\nu_I)
		[(1-\epsilon_I)\mathcal{H}_0]^{-1+2s-2\nu_I}}
	{16\,\Gamma(1-s+\nu_I)\Gamma(1-s+2\nu_I)\mathcal{H}^2}
	+ \frac{\mu_0^{-1+2s-2\nu_I}}{16\mathcal{H}^2(-\frac{1}{2}+s-\nu_I)} ,
\end{align}
where we were careful to write down the leading order $\mu_0$-dependent term.
The contribution to energy density from $(0,\mu_0)$ interval is then
\begin{align}
\rho_q^{(0,\mu_0)} ={}& \frac{3\mathcal{H}_0^4}{16\pi^2a^4} 
	\left( \frac{\mathcal{H}_1}{\mathcal{H}_0} \right)^{2\nu_I+1}
	\frac{\xi (1-6\xi) (2-\epsilon_I) (1-\epsilon_I)^2 (\nu_I-\frac{1}{2})^2}
		{(\nu_I-\frac{3}{2})}
\nonumber \\
&	- \frac{\mu_0^{3-2\nu_I}}{(\nu_I-\frac{3}{2})} \times 
	\frac{6\xi\Gamma^2(\nu_I)(\nu_I-\frac{1}{2})^2}{32\pi^3} 
	[2(1-\epsilon_I)\mathcal{H}_1]^{2\nu_I+1} ,
\end{align}
where we have kept the term that is $\mu_0$-independent in the
$\nu_I\rightarrow3/2$ limit. This contribution contributes the dominant term 
for $\xi<0$, and vanishes in the $\xi\rightarrow0$, $\epsilon_I\rightarrow0$,
$\mathcal{H}_0\rightarrow0$ limit.

On the interval $(\mu_0,\mu)$ we can again expand the Bogolyubov 
coefficients in $k/\mathcal{H}_1$, which amounts to the
integrand being approximated by
\begin{equation}
Z_{Bog.} = \frac{4|A_{2,1}|^2}{k^{2\nu_I+1}} |\alpha_{1,0}-\beta_{1,0}|^2
	\Re^2[u_R(k,\eta)] .
\end{equation}
From (83)!! we have that
\begin{equation}
|\alpha_{1,0}-\beta_{1,0}|^2 = \frac{2}{k} \left[\Re^2[u_I'(k,\eta_0)]
	+ k^2 \Re^2[u_I(k,\eta_0)] \right] ,
\end{equation}
which can be expanded in $\mathcal{H}_0/k$ on this interval,
\begin{equation}
|\alpha_{1,0}-\beta_{1,0}|^2 = 1 .
\end{equation}
Then we have for the $\mathcal{I}_s$ integrals (expanded in $\mu$ and $\mu_0$)
\begin{align}
\mathcal{I}_s^{(\mu_0,\mu)} ={}& 2|A_{2,1}|^2
	\int\limits_{\mu_0}^{\mu}\! dk\, k^{2s-2-2\nu_I} 
	\sin^2\left( \frac{k}{\mathcal{H}} \right)
\nonumber \\
={}& 2|A_{2,1}|^2  \left[ -4^{-s+\nu_I} \Gamma(-1+2s-2\nu_I)
	\sin[\pi(s-\nu_I)] + \frac{\mu^{-1+2s-2\nu_I}}{2(-1+2s-2\nu_I)} \right] .
\end{align}
Away from the limit in \cite{Glavan:2013mra} this interval gives a subdominant 
contribution to the energy density. Therefore, here we present just that particular
limit, which turns out to be
\begin{align}
\rho_q^{(\mu_0,\mu)} ={}& \frac{\mathcal{H}_1^4}{8\pi^2a^4} \ln(a) ,
\end{align}
which is exactly the one from \cite{Glavan:2013mra}. Actually, one has to go
through similar difficulties in the cases where $\nu_I$ or $\nu_M$ happen to 
have a half-integer value.

\section{Limits $\mathcal{H}\gg\mathcal{H}_0$ and
$\mathcal{H}\ll\mathcal{H}_0$ of the result during matter era}

In this appendix we present the direct computation of the limits
$\mathcal{H}\gg\mathcal{H}_0$ (early matter period) and
$\mathcal{H}\ll\mathcal{H}_0$ (very late matter period).
Even though the former physically makes sense in the limited range 
of non-minimal coupling, and the latter  not at all, it is still useful as 
an independent check of the results presented in Section IX.
Instead of calculating the $\mathcal{J}_s$ integrals, we calculate the
$\mathcal{I}_s$ integrals (\ref{Is IR}) straight away.

\subsection{Limit $\mathcal{H}\ll\mathcal{H}_0$}

In this case we solve the $\mathcal{I}_s$ integrals (\ref{Is IR}) with the hierarchy
$\mathcal{H}\ll\mu\ll\mathcal{H}_0\ll\mathcal{H}_2\ll\mathcal{H}_1$.
Here we can use the approximation of Section \ref{sec:Sudden transition IR}
for the entire Bogolyubov coefficients, namely
\begin{equation}
\alpha_{3,0} \approx \beta_{3,0} \approx \frac{iA_{3,0}}{k^{1/2+\nu_M}}
	\approx -i\beta_{3,0}^* ,
\end{equation}
where
\begin{equation}
A_{3,0} = \frac{\Gamma(\nu_M)(2-\epsilon_I)}
	{8\sqrt{\pi}\, \nu_I (1-\epsilon_I)}
	(1-6\xi) \left( \nu_M+\frac{3}{2} \right)
	\mathcal{H}_0^{-\nu_I+1/2}\mathcal{H}_1^{\nu_I+1/2}
	\mathcal{H}_2^{\nu_M-1/2} .
\end{equation}
This allows us to write the $\mathcal{I}_s$ integrals as
\begin{align}
\mathcal{I}_s ={}& 4|A_{3,0}|^2 \int\limits_{0}^{\mu}\! dk\, k^{2s-1-2\nu_M}
	\Re^2[u_M(k,\eta)]
	= \frac{2\pi|A_{3,0}|^2}{\mathcal{H}} \int\limits_{0}^{\mu}\! dk\, 
	k^{2s-1-2\nu_M} J_{\nu_M}^2 \left( \frac{2k}{\mathcal{H}} \right) ,
\end{align}
which evaluates to 
\begin{align}
\mathcal{I}_s ={}& |A_{3,0}|^2 \frac{\pi \mu^{2s}\mathcal{H}^{-1-2\nu_M}}
	{s\, \Gamma^2(1+\nu_M)}
	{}_2F_3\left( s, \frac{1}{2}+\nu_M; 1+s, 1+\nu_M, 1+2\nu_M;
	-\frac{4\mu^2}{\mathcal{H}^2} \right)
\nonumber \\
={}& |A_{3,0}|^2 \frac{2^{-2s+2\nu_M}\sqrt{\pi}\, \Gamma(s) 
	\Gamma(\frac{1}{2}-s+\nu_M)}{\Gamma(1-s+\nu_M) \Gamma(1-s+2\nu_M)}
	\, \mathcal{H}^{-1+2s-2\nu_M}
\end{align}
Plugging this into (\ref{rho H}) and (\ref{p H}) gives the very late time limit for the
energy density and pressure,
\begin{align}
\rho_q ={}& - \frac{(1-6\xi)^3(2-\epsilon_I)^2(\nu_M+\frac{3}{2})^2(\nu_M-1)
		(\nu_M-\frac{5}{2})}
	{2^9\pi^2(1-\epsilon_I)^2\nu_I^2(\nu_M-\frac{1}{2})(\nu_M-\frac{3}{2})}
 \times \mathcal{H}_0^{-2\nu_I+1}
	\mathcal{H}^{2\nu_I+1} \mathcal{H}_2^{2\nu_M-1} 
	\frac{\mathcal{H}^{3-2\nu_M}}{a^4} ,
 \\
p_q ={}& - \frac{(1-6\xi)^3(2-\epsilon_I)^2(\nu_M+\frac{3}{2})^2(\nu_M-1)
		(\nu_M-\frac{5}{2})}
	{2^9\pi^2(1-\epsilon_I)^2\nu_I^2(\nu_M-\frac{1}{2})(\nu_M-\frac{3}{2})}
	\times \frac{( \nu_M - \frac{5}{2} )}{3}
\nonumber \\
&	\qquad\qquad  \times \mathcal{H}_0^{-2\nu_I+1}
	\mathcal{H}^{2\nu_I+1} \mathcal{H}_2^{2\nu_M-1} 
	\frac{\mathcal{H}^{3-2\nu_M}}{a^4} .
\end{align}
This is precisely the limit one gets from expanding (\ref{J late}) in
$\mathcal{H}/\mathcal{H}_0\ll1$ for and then using it to calculate
the energy density and pressure. In this limit the backreaction behaves as 
a fluid with a constant equation of state,
\begin{equation}
w_q = \frac{1}{3}\left( \nu_M-\frac{5}{2} \right) ,
\end{equation}
and the sign of the energy density is determined by the sign of $-w_q$. 
The potentially interesting case where $\nu_M>5/2$ and the backreaction grows
with respect to the backrgound fluid corresponds to $\xi<-1/3$ which 
is far out of the region where the backreaction stays small during the evolution
of the Universe. In fact, it already becomes dominant in inflation. In the end,
this limit does not tell us much other than provides a check for the calculation
in Sections \ref{sec:Calculating IR} and \ref{sec:Results}.

Generally, the late time scaling in the limit $\mathcal{H}\ll\mathcal{H}_i$, where
$\mathcal{H}_i$ refer to the Hubble rates at any number $n$ of transitions
will depend only on the starting and ending deceleration periods as
$\sim a^{-4}\mathcal{H}^{4-2\nu_0-2\nu_n}$. That follows from the 
IR leading order term in the Bogolyubov coefficients derived in Section 
\ref{sec:Sudden transition IR} where it was found that
$\alpha\approx\beta\approx iA_{n,0} k^{-\nu_0-\nu_n}$.
The amplitude, of course, depends on the number and type of transitions.

\subsection{Limit $\mathcal{H}\gg\mathcal{H}_0$}

This limit corresponds to the early matter era, (up until today for minimal
inflation).
In this limit we can use approximation (\ref{Z approx1}) for the integrand
of $\mathcal{I}_s$ in (\ref{Is IR}),
\begin{align}
\mathcal{I}_s^{(0,\mu)} ={}& 
	4|A_{3,1}|^2 \int\limits_{0}^{\mu_0}\! dk\, k^{2s-2\nu_I-2\nu_M}
	|\alpha_{1,0}-\beta_{1,0}|^2 \Re^2[u_M(k,\eta)] ,
\end{align}
where
\begin{equation}
|\alpha_{1,0}-\beta_{1,0}|^2 = 
	\frac{2}{k} \left[ \Re^2[u_I'(k,\eta_0)] + k^2\Re[u_I(k,\eta_0)] \right] .
\end{equation}
Furthermore, we split the integration by
another scale $\mu_0$, where the hierarchy is
\begin{equation}
\mathcal{H}_0 \ll \mu_0 \ll \mathcal{H} \ll \mu \ll \mathcal{H}_2 
	\ll \mathcal{H}_1 .
\end{equation}
In the first of the two
integrals we can expand in $k/\mathcal{H}$. Performing this expansion
and using (\ref{84}) leads to
\begin{align}
\mathcal{I}_s^{(0,\mu_0)} = 8|A_{3,1}|^2 
	\left\{ 2\mathcal{J}_{s+1}^{(0,\mu_0)}
	+\left[ \frac{1}{2}\frac{\partial^2}{\partial\eta_0^2}+f_I(\eta_0) \right]
	\mathcal{J}_s^{(0,\mu_0)} \right\} ,
\end{align}
where
\begin{align}
\mathcal{J}_s^{(0,\mu_0)} ={}& \frac{\pi^2\mathcal{H}^{-1-2\nu_M}}
	{8(1-\epsilon_I)\Gamma^2(1+\nu_M)\mathcal{H}_0}
	\int\limits_{0}^{\mu_0}\! dk\, k^{2s-1-2\nu_I}
	J_{\nu_I}^2 \left( \frac{k}{(1-\epsilon_I)\mathcal{H}_0} \right)
\nonumber \\
={}& \frac{2^{-4-2\nu_I}\pi^2 \mathcal{H}^{-1-2\nu_M}\mu_0^{2s}
	[(1-\epsilon_I)\mathcal{H}_0]^{-1-2\nu_I}}
	{s\, \Gamma^2(1+\nu_I) \Gamma^2(1+\nu_M)}
\nonumber \\
&	\qquad \times {}_2F_3 \left( 2, \frac{1}{2}+\nu_I; 1+s, 1+\nu_I, 1+2\nu_I;
	\frac{-\mu_0^2}{(1-\epsilon_I)^2\mathcal{H}_0^2} \right)
\nonumber \\
={}& \frac{\pi^{3/2} \, \Gamma(s) \Gamma(\frac{1}{2}-s+\nu_I)}
	{16 \Gamma(1-s+\nu_I)\Gamma(1-s+2\nu_I)\Gamma^2(1+\nu_M)}
	[(1-\epsilon_I)\mathcal{H}_0]^{-1+2s-2\nu_I} \mathcal{H}^{-1-2\nu_I} .
\end{align}
This is precicely the leading order term in the limit $\mathcal{H}\gg\mathcal{H}_0$
of the full result (\ref{J early}) so here we must reproduce the same limit for 
energy density and pressure. So we have for the $\mu_0$-independent
 contribution to the 
$\mathcal{I}_s$ integrals
\begin{align} \label{Iless}
\mathcal{I}_s^{(0,\mu_0)} ={}& 8|A_{3,1}|^2 
	\frac{2^{-3+2s-2\nu_I} \pi^2 (1-6\xi)
	(2-\epsilon_I)(2\nu_I+1-2s)\Gamma(s)\Gamma(-1+2\nu_I-2s)}
	{(1-\epsilon_I)^2\Gamma(\nu_I-s)\Gamma(1-s+\nu_I)\Gamma(1-s+2\nu_I)
	\Gamma^2(1+\nu_M)} 
\nonumber \\
&	\qquad \times[(1-\epsilon_I)\mathcal{H}_0]^{1+2s-2\nu_I}
	\mathcal{H}^{-1-2\nu_I} .
\end{align}
On the other interval $(\mu_0,\mu)$ we can expand in $\mathcal{H}_0/k$, where
\begin{equation}
|\alpha_{1,0}-\beta_{1,0}|^2 = 1
\end{equation}
to leading order. This leads to
\begin{align}
\mathcal{I}_s^{(\mu_0,\mu)} ={}& \frac{2\pi|A_{3,1}|^2}{\mathcal{H}}
	\int\limits_{\mu_0}^{\mu}\! dk\, k^{2s-2\nu_I-2\nu_M}
	J_{\mu_M}^2 \left( \frac{2k}{\mathcal{H}} \right)
\nonumber \\
={}& |A_{3,1}|^2 \frac{2^{-1-2s+2\nu_I+2\nu_M}\sqrt{\pi}\,
	\Gamma(\frac{1}{2}+s-\nu_I) \Gamma(-s+\nu_I+\nu_M)}
	{\Gamma(\frac{1}{2}-s+\nu_I+\nu_M) 
	\Gamma(\frac{1}{2}-s+\nu_I+2\nu_M)} \mathcal{H}^{2s-2\nu_I-2\nu_M} ,
\end{align}
to leading order for the $\mu$-independent terms. This contribution is suppressed
compared to (\ref{Iless}). Therefore, the leading contribution to energy density 
and pressure during early matter era is
\begin{align}
\rho_q ={}& -\frac{(1-6\xi)(1-\epsilon_I)^2(2-\epsilon_I)(\nu_I-\frac{1}{2})^2
	(\nu_M+\frac{3}{2})^2(\nu_M-\frac{3}{2})(\nu_M+\frac{1}{2})}
	{2^8\pi^2\nu_M(\nu_I-\frac{3}{2})} 
\nonumber \\
&	\qquad \times \mathcal{H}_0^{3-2\nu_I} \mathcal{H}_1^{1+2\nu_I} 
	\mathcal{H}_2^{-1+2\nu_M} \mathcal{H}^{1-2\nu_M} a^{-4}
\nonumber \\
p_q ={}& -\frac{(1-6\xi)(1-\epsilon_I)^2(2-\epsilon_I)(\nu_I-\frac{1}{2})^2
	(\nu_M+\frac{3}{2})^2(\nu_M-\frac{3}{2})(\nu_M+\frac{1}{2})}
	{2^8\pi^2\nu_M(\nu_I-\frac{3}{2})} \times \frac{(\frac{3}{2}-\nu_M)}{3}
\nonumber \\
&	\qquad \times \mathcal{H}_0^{3-2\nu_I} \mathcal{H}_1^{1+2\nu_I} 
	\mathcal{H}_2^{-1+2\nu_M} \mathcal{H}^{1-2\nu_M} a^{-4} .
\end{align}
This result corresponds to the main result found in \cite{Glavan:2013mra}
for the limit $\epsilon_I\rightarrow0$, $\xi\rightarrow0$, 
$\mathcal{H}_0\rightarrow0$.


\begin{thebibliography}{99}

\bibitem{Glavan:2013mra}
  D.~Glavan, T.~Prokopec and V.~Prymidis,
  ``Backreaction of a massless minimally coupled scalar field from inflationary quantum fluctuations,''
  Phys.\ Rev.\ D {\bf 89} (2014) 024024
  [arXiv:1308.5954 [gr-qc]].


\bibitem{Aoki:2014ita}
  H.~Aoki, S.~Iso and Y.~Sekino,
  ``Evolution of vacuum fluctuations generated during and before inflation,''
  Phys.\ Rev.\ D {\bf 89} (2014) 103536
  [arXiv:1402.6900 [hep-th]].


\bibitem{Ade:2014xna}
  P.~A.~R.~Ade {\it et al.}  [BICEP2 Collaboration],
  ``Detection of B-Mode Polarization at Degree Angular Scales by BICEP2,''
  Phys.\ Rev.\ Lett.\  {\bf 112} (2014) 241101
  [arXiv:1403.3985 [astro-ph.CO]].
  

\bibitem{Kolb:2005da}
  E.~W.~Kolb, S.~Matarrese and A.~Riotto,
  ``On cosmic acceleration without dark energy,''
  New J.\ Phys.\  {\bf 8} (2006) 322
  [astro-ph/0506534].


\bibitem{Barausse:2005nf}
  E.~Barausse, S.~Matarrese and A.~Riotto,
  ``The Effect of inhomogeneities on the luminosity distance-redshift relation: Is dark energy necessary in a perturbed Universe?,''
  Phys.\ Rev.\ D {\bf 71} (2005) 063537
  [astro-ph/0501152].


\bibitem{Hirata:2005ei}
  C.~M.~Hirata and U.~Seljak,
  ``Can superhorizon cosmological perturbations explain the acceleration of the Universe?,''
  Phys.\ Rev.\ D {\bf 72} (2005) 083501
  [astro-ph/0503582].


\bibitem{Janssen:2009nz}
  T.~M.~Janssen and T.~Prokopec,
  ``Regulating the infrared by mode matching: A Massless scalar in expanding spaces with constant deceleration,''
  Phys.\ Rev.\ D {\bf 83} (2011) 084035
  [arXiv:0906.0666 [gr-qc]].


\bibitem{Abramo:2001dc}
  L.~R.~Abramo and R.~P.~Woodard,
  ``No one loop back reaction in chaotic inflation,''
  Phys.\ Rev.\ D {\bf 65} (2002) 063515
  [astro-ph/0109272].


\bibitem{Abramo:2001db}
  L.~R.~Abramo and R.~P.~Woodard,
  ``A Scalar measure of the local expansion rate,''
  Phys.\ Rev.\ D {\bf 65} (2002) 043507
  [astro-ph/0109271].


\bibitem{Abramo:1998hj}
  L.~R.~W.~Abramo and R.~P.~Woodard,
  ``One loop back reaction on power law inflation,''
  Phys.\ Rev.\ D {\bf 60} (1999) 044011
  [astro-ph/9811431].


\bibitem{Abramo:1998hi}
  L.~R.~W.~Abramo and R.~P.~Woodard,
  ``One loop back reaction on chaotic inflation,''
  Phys.\ Rev.\ D {\bf 60} (1999) 044010
  [astro-ph/9811430].


\bibitem{Geshnizjani:2002wp}
  G.~Geshnizjani and R.~Brandenberger,
  ``Back reaction and local cosmological expansion rate,''
  Phys.\ Rev.\ D {\bf 66} (2002) 123507
  [gr-qc/0204074].


\bibitem{Geshnizjani:2003cn}
  G.~Geshnizjani and R.~Brandenberger,
  ``Back reaction of perturbations in two scalar field inflationary models,''
  JCAP {\bf 0504} (2005) 006
  [hep-th/0310265].


\bibitem{Abramo:2001dd}
  L.~R.~Abramo and R.~P.~Woodard,
  ``Back reaction is for real,''
  Phys.\ Rev.\ D {\bf 65} (2002) 063516
  [astro-ph/0109273].


\bibitem{Janssen:2008dw}
  T.~Janssen and T.~Prokopec,
  ``The Graviton one-loop effective action in cosmological space-times with constant deceleration,''
  Annals Phys.\  {\bf 325} (2010) 948
  [arXiv:0807.0447 [gr-qc]].


\bibitem{Marozzi:2013uva}
  G.~Marozzi and G.~P.~Vacca,
  ``Gauge invariant backreaction in general single field models of inflation,''
  Phys.\ Rev.\ D {\bf 88} (2013) 027302
  [arXiv:1304.2291 [gr-qc]].


\bibitem{Marozzi:2011zb}
  G.~Marozzi and G.~P.~Vacca,
  ``Isotropic Observers and the Inflationary Backreaction Problem,''
  Class.\ Quant.\ Grav.\  {\bf 29} (2012) 115007
  [arXiv:1108.1363 [gr-qc]].


\bibitem{Marozzi:2012tp}
  G.~Marozzi, G.~P.~Vacca and R.~H.~Brandenberger,
  ``Cosmological Backreaction for a Test Field Observer in a Chaotic Inflationary Model,''
  JCAP {\bf 1302} (2013) 027
  [arXiv:1212.6029 [hep-th]].


\bibitem{Prokopec:2006ue}
  T.~Prokopec, N.~C.~Tsamis and R.~P.~Woodard,
  ``Two Loop Scalar Bilinears for Inflationary SQED,''
  Class.\ Quant.\ Grav.\  {\bf 24} (2007) 201
  [gr-qc/0607094].


\bibitem{Prokopec:2008gw}
  T.~Prokopec, N.~C.~Tsamis and R.~P.~Woodard,
  ``Two loop stress-energy tensor for inflationary scalar electrodynamics,''
  Phys.\ Rev.\ D {\bf 78} (2008) 043523
  [arXiv:0802.3673 [gr-qc]].


\bibitem{Prokopec:2007ak}
  T.~Prokopec, N.~C.~Tsamis and R.~P.~Woodard,
  ``Stochastic Inflationary Scalar Electrodynamics,''
  Annals Phys.\  {\bf 323} (2008) 1324
  [arXiv:0707.0847 [gr-qc]].


\bibitem{Miao:2006pn}
  S.~-P.~Miao and R.~P.~Woodard,
  ``Leading log solution for inflationary Yukawa,''
  Phys.\ Rev.\ D {\bf 74} (2006) 044019
  [gr-qc/0602110].


\bibitem{Lazzari:2013boa}
  G.~Lazzari and T.~Prokopec,
  ``Symmetry breaking in de Sitter: a stochastic effective theory approach,''
  arXiv:1304.0404 [hep-th].


\bibitem{Serreau:2013eoa}
  J.~Serreau,
  ``Renormalization group flow and symmetry restoration in de Sitter space,''
  Phys.\ Lett.\ B {\bf 730} (2014) 271
  [arXiv:1306.3846 [hep-th]].


\bibitem{Prokopec:2011ms}
  T.~Prokopec,
  ``Symmetry breaking and the Goldstone theorem in de Sitter space,''
  JCAP {\bf 1212} (2012) 023
  [arXiv:1110.3187 [gr-qc]].


\bibitem{Serreau:2011fu}
  J.~Serreau,
  ``Effective potential for quantum scalar fields on a de Sitter geometry,''
  Phys.\ Rev.\ Lett.\  {\bf 107} (2011) 191103
  [arXiv:1105.4539 [hep-th]].


\bibitem{Janssen:2009pb}
  T.~M.~Janssen, S.~P.~Miao, T.~Prokopec and R.~P.~Woodard,
  ``The Hubble Effective Potential,''
  JCAP {\bf 0905} (2009) 003
  [arXiv:0904.1151 [gr-qc]].
  
  
\bibitem{Starobinsky:1994bd}
  A.~A.~Starobinsky and J.~Yokoyama,
  ``Equilibrium state of a selfinteracting scalar field in the De Sitter background,''
  Phys.\ Rev.\ D {\bf 50} (1994) 6357
  [astro-ph/9407016].

  
\bibitem{Vilenkin:1982wt}
  A.~Vilenkin and L.~H.~Ford,
  ``Gravitational Effects upon Cosmological Phase Transitions,''
  Phys.\ Rev.\ D {\bf 26} (1982) 1231.


\bibitem{Ford:1977in}
  L.~H.~Ford and L.~Parker,
  ``Infrared Divergences in a Class of Robertson-Walker Universes,''
  Phys.\ Rev.\ D {\bf 16} (1977) 245.


\bibitem{Tsamis:2002qk}
  N.~C.~Tsamis and R.~P.~Woodard,
  ``Plane waves in a general Robertson-Walker background,''
  Class.\ Quant.\ Grav.\  {\bf 20} (2003) 5205
  [astro-ph/0206010].


\bibitem{Birrell:1982ix}
  N.~D.~Birrell and P.~C.~W.~Davies,
  ``Quantum Fields in Curved Space,''
  

\bibitem{Capper:1974ic}
  D.~M.~Capper and M.~J.~Duff,
  ``Trace anomalies in dimensional regularization,''
  Nuovo Cim.\ A {\bf 23} (1974) 173.
  

\bibitem{Duff:1977ay}
  M.~J.~Duff,
  ``Observations on Conformal Anomalies,''
  Nucl.\ Phys.\ B {\bf 125} (1977) 334.
  

\bibitem{Weinberg:2010wq}
  S.~Weinberg,
  ``Ultraviolet Divergences in Cosmological Correlations,''
  Phys.\ Rev.\ D {\bf 83} (2011) 063508
  [arXiv:1011.1630 [hep-th]].


\bibitem{Koivisto:2010pj}
  T.~S.~Koivisto and T.~Prokopec,
  ``Quantum backreaction in evolving FLRW spacetimes,''
  Phys.\ Rev.\ D {\bf 83} (2011) 044015
  [arXiv:1009.5510 [gr-qc]].


\bibitem{Albareti:2014ria}
  F.~D.~Albareti, J.~A.~R.~Cembranos and A.~L.~Maroto,
  ``Vacuum energy as dark matter,''
  arXiv:1404.5946 [gr-qc].


\bibitem{Sheikhahmadi:2014rka}
  H.~Sheikhahmadi, A.~Aghamohammadi and K.~Saaidi,
  ``Vacuum quantum fluctuations in a quasi de Sitter background,''
  arXiv:1407.0125 [gr-qc].


\bibitem{Abramo:2001dd}
  L.~R.~Abramo and R.~P.~Woodard,
  ``Back reaction is for real,''
  Phys.\ Rev.\ D {\bf 65} (2002) 063516
  [astro-ph/0109273].


\bibitem{Abramo:1998hi}
  L.~R.~W.~Abramo and R.~P.~Woodard,
  ``One loop back reaction on chaotic inflation,''
  Phys.\ Rev.\ D {\bf 60} (1999) 044010
  [astro-ph/9811430].


\bibitem{Kahya:2009sz}
  E.~O.~Kahya, V.~K.~Onemli and R.~P.~Woodard,
  ``A Completely Regular Quantum Stress Tensor with w < -1,''
  Phys.\ Rev.\ D {\bf 81} (2010) 023508
  [arXiv:0904.4811 [gr-qc]].


\bibitem{Tsamis:1996qm}
  N.~C.~Tsamis and R.~P.~Woodard,
  ``The Quantum gravitational back reaction on inflation,''
  Annals Phys.\  {\bf 253} (1997) 1
  [hep-ph/9602316].


\bibitem{Tsamis:1996qq}
  N.~C.~Tsamis and R.~P.~Woodard,
  ``Quantum gravity slows inflation,''
  Nucl.\ Phys.\ B {\bf 474} (1996) 235
  [hep-ph/9602315].


\bibitem{Ringeval:2010hf}
  C.~Ringeval, T.~Suyama, T.~Takahashi, M.~Yamaguchi and S.~Yokoyama,
  ``Dark energy from primordial inflationary quantum fluctuations,''
  Phys.\ Rev.\ Lett.\  {\bf 105} (2010) 121301
  [arXiv:1006.0368 [astro-ph.CO]].


\bibitem{Suen:1987gu}
  W.~M.~Suen and P.~R.~Anderson,
  ``Reheating in the Higher Derivative Inflationary Models,''
  Phys.\ Rev.\ D {\bf 35} (1987) 2940.


\bibitem{Koksma:2008jn}
  J.~F.~Koksma and T.~Prokopec,
  ``The Effect of the Trace Anomaly on the Cosmological Constant,''
  Phys.\ Rev.\ D {\bf 78} (2008) 023508
  [arXiv:0803.4000 [gr-qc]].


\bibitem{Janssen:2007ht}
  T.~Janssen and T.~Prokopec,
  ``A Graviton propagator for inflation,''
  Class.\ Quant.\ Grav.\  {\bf 25} (2008) 055007
  [arXiv:0707.3919 [gr-qc]].


\bibitem{Prokopec:2002jn}
  T.~Prokopec, O.~Tornkvist and R.~P.~Woodard,
  ``Photon mass from inflation,''
  Phys.\ Rev.\ Lett.\  {\bf 89} (2002) 101301
  [astro-ph/0205331].


\bibitem{Prokopec:2002uw}
  T.~Prokopec, O.~Tornkvist and R.~P.~Woodard,
  ``One loop vacuum polarization in a locally de Sitter background,''
  Annals Phys.\  {\bf 303} (2003) 251
  [gr-qc/0205130].


\bibitem{Ade:2013ktc}
  P.~A.~R.~Ade {\it et al.}  [Planck Collaboration],
  ``Planck 2013 results. I. Overview of products and scientific results,''
  arXiv:1303.5062 [astro-ph.CO].





\end{thebibliography}
\end{document}